\documentclass[dvips,longauth,traditabstract]{aa} 
\usepackage{mathtools}
\usepackage{graphicx}
\usepackage{txfonts}
\usepackage{natbib}
\usepackage{longtable}

\def\msol{$M_{\odot}$}
\def\micron{$\mu$m}

\def\nhtwo{$N_{\rm H_2}$}

\def\alphabs{$\bar{\alpha}_{\rm corr} =-1.6\pm0.3$}
\def\alphanbs{$\bar{\alpha}_{\rm obs} = -2.0\pm0.2$}
\def\alphafilbs{$-1.6\pm0.2$}
\def\alphafil{$-2.1\pm0.2$}
\def\ampfilbs{9.5$\times10^{-2}$ \msol$^2$/pc} 
\def\ampfil{6$\times10^{-2}$ \msol$^2$/pc}

\def\pmline{$1.4\pm0.1$}

\def\her{\emph{Herschel}}

\begin{document}
\title{A possible link between the power spectrum of interstellar filaments and the origin of
the prestellar core mass function}

\author{A. Roy\inst{\ref{inst1}}
  \and Ph. Andr\'{e}\inst{\ref{inst1}}
  \and D. Arzoumanian \inst{\ref{inst1},\ref{inst-doris}}
  \and N. Peretto \inst{\ref{inst-cardif}}
  \and P. Palmeirim\inst{\ref{inst1}} 
  \and V. K\"{o}nyves\inst{\ref{inst1}}
   \and N. Schneider \inst{\ref{inst1},\ref{inst-nicola-s}}
  \and M. Benedettini \inst{\ref{inst-INAF}}
  \and J. Di Francesco \inst{\ref{inst-uvic}, \ref{inst-nrcc}}   
  \and D. Elia\inst{\ref{inst-INAF}}
  \and T. Hill \inst{\ref{inst1},\ref{inst-alma}}
  \and B. Ladjelate \inst{\ref{inst1}}
  \and F. Louvet\inst{\ref{inst-chile} }
  \and F. Motte\inst{\ref{inst1}}
  \and S. Pezzuto \inst{\ref{inst-INAF}}
  \and E. Schisano \inst{\ref{inst-INAF}}
   \and Y. Shimajiri\inst{\ref{inst1}}
  \and L. Spinoglio \inst{\ref{inst-INAF}}
  \and D. Ward-Thompson\inst{\ref{inst-lanc}}
   \and G. White \inst{\ref{inst-white1},\ref{inst-white2}}
    }

\institute{Laboratoire AIM, CEA/DSM-CNRS-Universit\'{e} Paris Diderot,
IRFU / Service d'Astrophysique, C.E. Saclay, Orme des
Merisiers, 91191 Gif-sur-Yvette\label{inst1} 
\and Institut d'Astrophysique Spatiale, CNRS/Universit\'{e} Paris-Sud 11, 91405 Orsay, France \label{inst-doris}
\and  School of Physics \& Astronomy, Cardiff University, Cardiff, CF29, 3AA, UK \label{inst-cardif}
\and Universit\'{e} de Bordeaux, Laboratoire d’Astrophysique de Bordeaux, CNRS/INSU, UMR 5804, BP 89, 33271, Floirac Cedex, France \label{inst-nicola-s}
\and INAF-Istituto di Astrofisica e Planetologia Spaziali, via Fosso del Cavaliere 100, I-00133 Rome, Italy \label{inst-INAF}
\and Department of Physics and Astronomy, University of Victoria, P.O. Box 355, STN CSC, Victoria, BC, V8W 3P6, Canada \label{inst-uvic}
\and National Research Council Canada, 5071 West Saanich Road, Victoria, BC, V9E 2E7, Canada \label{inst-nrcc}
\and Joint ALMA Observatory, Alonso de \'{C}ordova 3107, Vitacura, Santiago, Chile \label{inst-alma}
\and Departamento de Astronom\'{i}a, Universidad de Chile, Santiago, Chile  \label{inst-chile}
\and Jeremiah Horrocks Institute, University of Central Lancashire, Preston, Lancashire, PR1 2HE, UK \label{inst-lanc}
\and Department of Physics and Astronomy, The Open University, Walton Hall Milton Keynes, MK7 6AA, United Kingdom\label{inst-white1}
\and RAL Space, STFC Rutherford Appleton Laboratory, Chilton Didcot, Oxfordshire OX11 0QX, United Kingdom\label{inst-white2}\\
 E-mails: Arabindo.Roy@cea.fr, philippe.andre@cea.fr}

\titlerunning{Power spectra} \abstract{ A complete understanding of
  the origin of the prestellar core mass function (CMF) is
  crucial. Two major features of the prestellar CMF are: 1$)$ a broad
  peak below $1\, M_\odot $, presumably corresponding to a mean
  gravitational fragmentation scale, and 2$)$ a characteristic
  power-law slope, very similar to the Salpeter slope of the stellar
  initial mass function (IMF) at the high-mass end.  While recent
  \her\ observations have shown that the peak of the prestellar CMF is
  close to the thermal Jeans mass in marginally supercritical
  filaments, the origin of the power-law tail of the CMF/IMF at the
  high-mass end is less clear.  Inutsuka (2001) proposed a theoretical
  scenario in which the origin of the power-law tail can be understood
  as resulting from the growth of an initial spectrum of density
  perturbations seeded along the long axis of star-forming filaments
  by interstellar turbulence.  Here, we report the statistical
  properties of the line-mass fluctuations of filaments in the Pipe,
  Taurus, and IC5146 molecular clouds observed with {\it Herschel} for
  a sample of subcritical or marginally supercritical filaments using
  a 1-D power spectrum analysis.  The observed filament power spectra
  were fitted by a power-law function ($P_{\rm true}(s) \propto
  s^{\alpha}$) after removing the effect of beam convolution at small
  scales.  A Gaussian-like distribution of power-spectrum slopes was
  found, centered at \alphabs.  The characteristic index of the
  observed power spectra is close to that of the one-dimensional
  velocity power spectrum generated by subsonic Kolomogorov turbulence
  ($-1.67$).  Given the errors, the measured power-spectrum slope is
  also marginally consistent with the power spectrum index of $-2$
  expected for supersonic compressible turbulence.  With such a power
  spectrum of initial line-mass fluctuations, Inutsuka's model would
  yield a mass function of collapsed objects along filaments
  approaching $dN/dM \propto M^{-2.3\pm0.1}$ at the high-mass end
  (very close to the Salpeter power law) after a few free-fall times.
  An empirical correlation, $P^{0.5}(s_0) \propto \langle N_{\rm H_2}
  \rangle^{1.4 \pm 0.1} $, was also found between the amplitude of
  each filament power spectrum $P(s_0)$ and the mean column density
  along the filament $ \langle N_{\rm H_2} \rangle $.  Finally, the
  dispersion of line-mass fluctuations along each filament
  $\sigma_{\rm M_{line}}$ was found to scale with the physical length
  $L$ of the filament, roughly as $\sigma_{\rm M_{line}} \propto
  L^{0.7}$.  Overall, our results are consistent with the suggestion
  that the bulk of the CMF/IMF results from the gravitational
  fragmentation of filaments.  }

\maketitle

\section{Introduction}

Understanding the origin of the stellar initial mass function (IMF) is
a fundamental open problem in modern astrophysics
(e.g. \citealp{offner2013} for a recent review).  Since the end of the
1990s, several observational studies of prestellar dense cores in
nearby molecular clouds have found a strong link between the
prestellar core mass function (CMF) and the IMF (e.g., see
\citealp{motte1998, alves2007, konyves2010, konyves2015}), suggesting
that the IMF is at least partly the result of the core formation
process.  Theories of the CMF/IMF based on gravo-turbulent
fragmentation (e.g. \citealp{Padoan2002, hennebelle2008},
\citealp{hopkins2012}) are consistent with this view but do not
account for the fact that most cores/stars appear to form within
interstellar filaments.  Indeed, recent {\it Herschel} observations
(e.g., \citealp{andre2010, molinari2010, arzoumanian2011, hill2011,
  konyves2015}) emphasize the role of interstellar filaments in
star-forming clouds and support a paradigm for star formation in which
the formation of $\sim 0.1$~pc-wide filaments and the subsequent
fragmentation of the densest filaments into prestellar cores represent
two key steps in the star formation process
(cf. \citealp{Andre+2014}).  The results of the {\it Herschel} Gould
Belt survey (HGBS) further suggest that gravitational fragmentation of
marginally supercritical filaments with masses per unit length $M_{\rm
  line} $ approaching the critical line mass of nearly isothermal
$\sim 10$~K gas cylinders $M_{\rm line, crit }=2c^2_{\rm s}/G \sim
16$~\msol/pc (\citealp{ostriker1964, inutsuka1997}) may be responsible
for the peak of the prestellar CMF at $\sim 0.6$~\msol, as observed in
the Aquila cloud complex for example \citep{Andre+2014, konyves2015}.
Indeed, this idea is consistent with the view that the peak of the IMF
is related to the typical Jeans mass in star-forming clouds
(\citealp{larson1985}).

In gravo-turbulent fragmentation theories of the CMF/IMF, the peak of
the IMF results from a combination of thermal physics, setting the
mean thermal Jeans mass in the parent cloud, and turbulence effects,
through the turbulence Mach number \citep{hennebelle2008, hopkins2012,
  chabrier2014}.  In the filamentary picture proposed by
\citet{Andre+2014}, a characteristic thermal Jeans mass results from
the existence of a critical line mass for filaments (which depends
only on gas temperature) and from the characteristic filament width
$\sim 0.1$~pc measured with {\it Herschel} \citep{arzoumanian2011},
which is close to the sonic scale of turbulence in low-density
molecular gas.  Taken together, this sets a (column) density threshold
for prestellar core and star formation as observed in nearby clouds
with {\it Herschel} \citep{andre2010, konyves2015} and {\it Spitzer}
\citep{heiderman2010,lada2010}, respectively.

\begin{figure*}
  \centering
  \resizebox{\hsize}{!}{\includegraphics[angle=0]{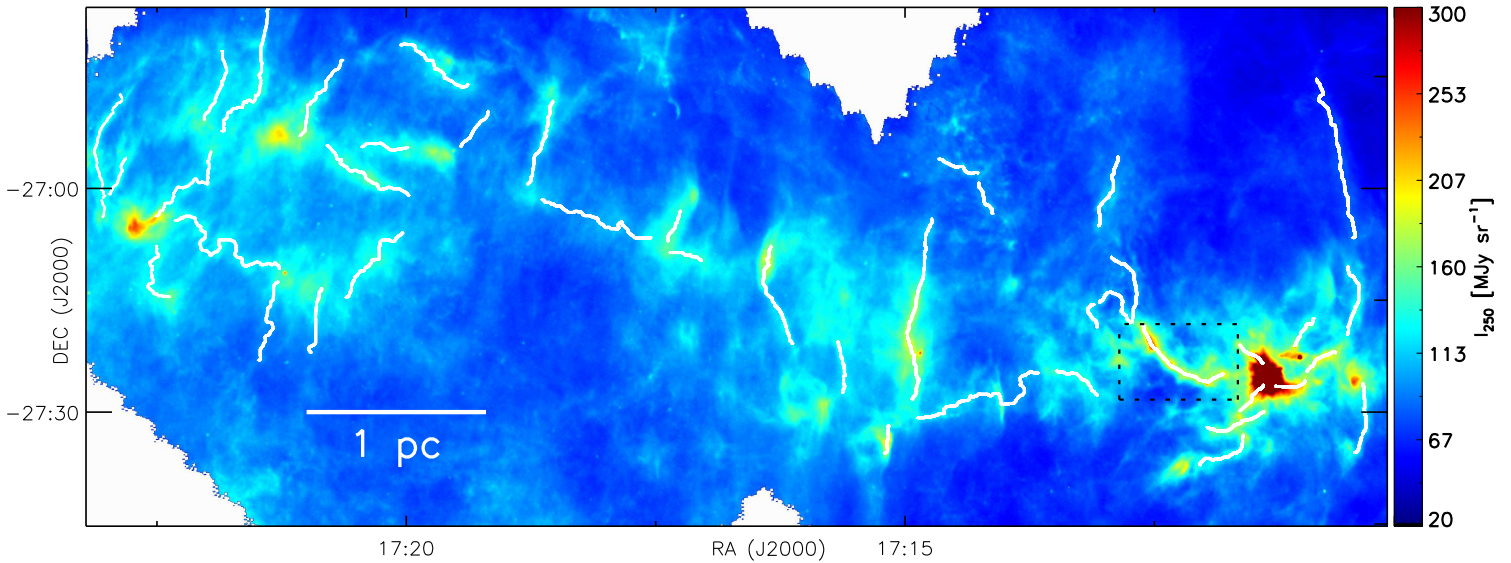}}
  \caption{\her\ 250~\micron\ surface brightness image of the Pipe
    molecular cloud observed as a part of the HGBS key project
    \citep[cf.][]{peretto2012}.  The image is displayed at the
    diffraction limited 250~\micron\ beam resolution of 18\farcs2. The
    43 filaments traced with the DisPerSE algorithm
    \citep{sousbie2011} are highlighted in white.  A more detailed
    view of the filament enclosed by the black dashed rectangle is
    shown in Fig.~\ref{fig:fluc}.}
\label{fig:pipefig}
\end{figure*}

Although the thermal Jeans mass within marginally supercritical filaments may account
for the peak of the prestellar CMF, pure thermal fragmentation of filaments at the threshold 
of gravitational instability is expected to lead to a sharply peaked CMF (such as a delta function)
and can hardly explain the Salpeter power-law regime of the IMF at the high-mass end. 
\cite{inutsuka2001} nevertheless suggested that 
a Salpeter-like CMF can quickly develop within filaments provided that turbulence has 
generated the appropriate power spectrum of initial density fluctuations in the first place. 
More specifically, 
\cite{inutsuka2001} found that the hierarchical structural property of
the perturbed density field along filaments may statistically produce
a population of cores with a power-law mass function.  
Inutsuka's approach
consisted of counting the distribution of ``isolated'' collapsed regions of mass scale
$M$ above a critical density using the Press-Schechter formalism \citep{press1974,jedamzik1995}, 
where the statistics of such regions depend on the logarithmic slope of
the power spectrum characterizing the initial fluctuating density field.  
Inutsuka (2001) showed that the resulting core mass distribution converges 
toward a Salpeter-like mass function $dN/dM \propto M^{-2.5}$ 
when the power spectrum of the field of initial line-mass
fluctuations approaches $P(s)\propto s^{-1.5}$. 
The latter power spectrum was adopted on a purely ad-hoc basis, however, 
and the true statistical nature of the density fluctuations along filaments remained 
to be quantified observationally. 
Another important assumption of Inutsuka's model was that the 
initial line-mass fluctuations along filaments could be represented by a Gaussian random distribution. 

In this paper, we exploit the unprecedented sensitivity and resolution of
\her\ submillimeter continuum images as well as the proximity of 
 the molecular clouds targeted as part of the HGBS key project
to characterize, for the first time, the statistical properties
of the line-mass fluctuations along interstellar filaments by analyzing
their 1-D power spectrum. Our main goal is to constrain observationally 
the validity of the theoretical model outlined by
\cite{inutsuka2001} for the origin of a Salpeter-like prestellar CMF above $\sim 1\, M_\odot $.

\section{Herschel observations and column density maps}\label{sec:obs}

The three target fields that we used in the present analysis 
were all observed with the {\it Herschel}\footnote{Herschel is an ESA space observatory 
with science instruments provided by European-led Principal Investigator consortia and with important participation from NASA.} 
space observatory (\citealp{pilbratt2010}) 
as part of the HGBS key project \citep{andre2010}, 
and cover surface areas of $\sim
1\rlap{.}\degr 5 \times 1\rlap{.}\degr5$, $\sim 6\degr \times
2\rlap{.}\degr5$, and $\sim 1\rlap{.}\degr 6 \times 1\rlap{.}\degr6$,
in the IC5146, Pipe, and Taurus molecular clouds, respectively
\citep{arzoumanian2011,peretto2012,palmeirim2013}.
These three regions were mapped at a scanning speed of
$60\arcsec \rm{s}^{-1}$ in parallel mode simultaneously at five
\her\ wavelengths using the SPIRE \citep{griffin2010} and PACS
\citep{poglitsch2010} photometric cameras.  The data were reduced using HIPE
version 7.0. For the SPIRE data reduction, we used modified pipeline
scripts. Observations during the turnaround of the telescope were
included, and a destriper module with a zero-order polynomial baseline
was applied. The default `na\"{\i}ve' mapper was used to produce the
final maps.  For the PACS data, we applied the standard HIPE data
reduction pipeline up to level 1, with improved calibration. Further
processing of the data, such as subtraction of (thermal and
non-thermal) low-frequency noise and map projection was performed with
Scanamorphos v11 \citep{roussel2013}. Note that the Scannamorphos
map-maker avoids any high-pass filtering, which is crucial for
preserving extended emission.
We adjusted the zero-point values of individual \her\ images based on cross-correlations 
of the \her\ data with IRAS and \emph{Planck} data \citep[cf.][]{bernard2010}. Figure \ref{fig:pipefig}
shows an example of 250~\micron\ surface brightness image of the Pipe Nebula at 
the native resolution of 18\farcs2.

\begin{figure}[!h]
  \centering
  \resizebox{0.95\hsize}{!}{\includegraphics[angle=0]{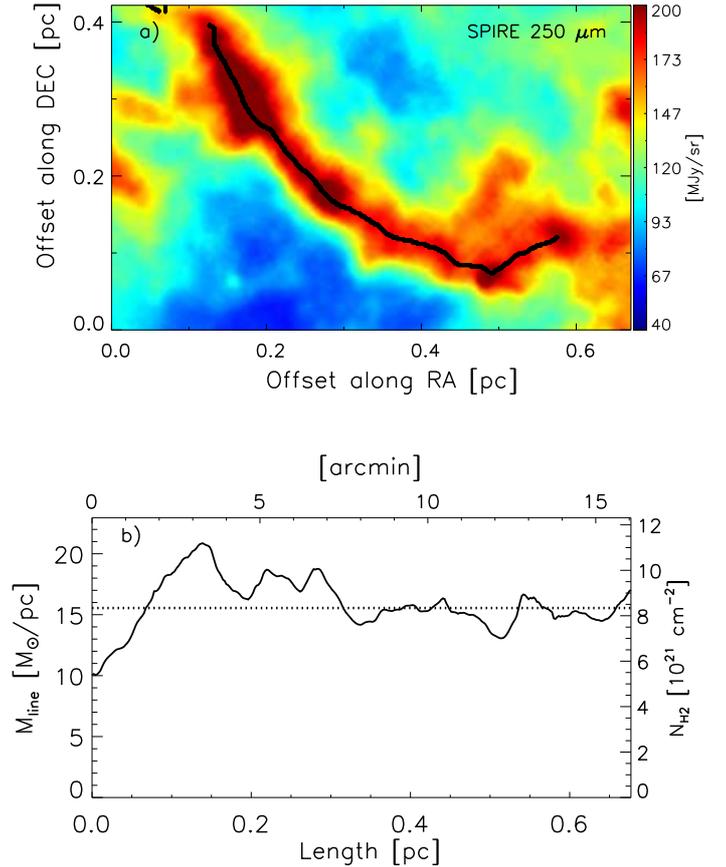}}
  \caption{Close-up view of the sub-region marked by a black dashed
    rectangle in Fig.~\ref{fig:pipefig}, showing ({\bf a$)$} an
    example of a nearly critical filament in the Pipe cloud at an
    adopted distance $d = 145$~pc, and ({\bf b$)$} the
    line-mass/column density fluctuations along the long axis of the
    same filament, both at the 18\farcs2 (HPBW) resolution of the
    SPIRE 250~\micron\ data.  The black curve shown in {(\bf a$)$}
    marks the crest of the filament as traced by the DisPerSE
    algorithm \citep[cf.][]{sousbie2011}.  In {\bf b$)$} the right
    ordinate axis shows the net H$_{2}$-column density fluctuations
    along the filament.  Equivalent line-mass values (assuming a
    characteristic filament width of 0.1~pc) are indicated on the left
    ordinate axis.  A constant background column density level was
    subtracted before deriving the line mass fluctuations.  The dotted
    horizontal line marks the mean line mass of the filament, $\langle
    M_{\rm line}\rangle \sim $ 15 \msol/pc, which is only slightly
    lower the critical line mass of 16.3 \msol/pc for an isothermal
    cylinder at 10~K.  The median relative amplitude of the line-mass
    fluctuations is $ \frac{ \lvert M_{\rm line}- \langle M_{\rm line}
      \rangle \rvert}{\langle M_{\rm line} \rangle} \sim 0.07$, within
    the linear perturbation regime. The maximum relative amplitude is
    $\sim$ 0.3 $\ll$ 1. }
\label{fig:fluc}
\end{figure}

For the present analysis, we needed high-resolution column density
maps for recovering small-scale fluctuations which would otherwise be
smeared due to beam convolution.  We employed two methods for
generating such column density maps.  First, we generated column
density maps at an effective resolution corresponding to the SPIRE
250~\micron\ beam following the method described in Appendix~A of
\cite{palmeirim2013}. In this method, large-scale information is
obtained from a `standard' {\it Herschel} column density map
constructed at the SPIRE 500~\micron\ resolution, while finer details
are recovered by constructing more approximate column density maps at
the resolution of SPIRE 350~\micron\ and SPIRE 250~\micron\ data,
respectively.  Following the steps described in Palmeirim et al., we
fitted a single-temperature modified blackbody to the observed data on
a pixel by pixel basis, adopting a wavelength-dependent dust opacity
of the form $\kappa_{\lambda} = 0.1 (\lambda/300~\mu \rm m)^{-\beta}$
cm$^{2}$ per g (of gas $+$ dust) with an emissivity index $\beta =2$
\citep[cf.][]{Hildebrand1983,Roy2014}.  In the second, alternative
method, we directly converted the surface brightness maps observed at
250~\micron\ $I_{\rm 250} $ into approximate column density maps by
using the relation \nhtwo =$I_{250}/(B_{250}[T_{\rm d, SED}]
\kappa_{250} \mu_{\rm H_2} m_{\rm H})$, where $\mu_{\rm H_2}=2.8$ is
the mean molecular weight and $T_{\rm d, SED}$ is the SED dust
temperature obtained by fitting a modified blackbody to the SED
observed between 160~\micron\ and 500~\micron\ with \her\ data toward
each line of sight.  For the Pipe Nebula filament shown in
Fig.~\ref{fig:fluc}, for instance, the median dust temperature along
the filament crest is $\bar{T}_{\rm d, SED} =13.5$~K.  The advantage
of the first method is that it produces a more accurate column density
map at the $18.2\arcsec $ resolution of the SPIRE
250~\micron\ data. The disadvantage, however, is that the resulting
map is significantly noisier than the SPIRE 250~\micron\ map and the
point spread function (PSF) is more difficult to characterize due to
the various steps involved in the processing.  In contrast, the second
method which simply produces modified 250 \micron\ maps, the effective
PSF is exactly the same as that of the SPIRE 250 \micron\ photometer.
Since an accurate knowledge of the beam is important in the present
power-spectrum study, we report results based on the latter method in
the main body of the paper.  The results of a power-spectrum analysis
performed using high-resolution column density maps produced with the
former method are given in Appendix~B. They confirm the robustness of
our conclusions.
 
\section{Filament identification}  \label{sec:fil_identification}

For the purpose of detecting coherently elongated filamentary
structures in the column density maps as well as the modified
250~\micron\ images, we employed the DisPerSE algorithm which traces
the crests of (segments of) ``topological'' filaments by connecting
saddle points to maxima following the gradient in an image
(\citealp{sousbie2011}; see also \citealp{arzoumanian2011,hill2011}
for details on practical applications of DisPerSE on {\it Herschel}
images).  It is important to point out here that DisPerSE was used
solely for tracing the filament crests, while the filament
longitudinal profiles were measured directly on the original images.
Our filament sample consists of 28 filaments in the Taurus cloud
(adopted distance $d = 140$~pc; \citealp{elias1978}), 42 filaments in
the Pipe nebula region ($d = 145$~pc; \citealp{alves2007}), and 36
filaments in the IC5146 cloud ($d = 460$~pc; \citealp{lada1999}).
Combining the three regions, we identified a total of 106 filaments of
various lengths above a 5 $\sigma$ persistence level of $\sim
10^{21}\, \rm{cm}^{-2} $.  For a proper power-spectrum analysis which
was our ultimate goal, however, we had to select moderately long
filaments to have an adequate number of Fourier modes at small spatial
frequencies (see Sect.~\ref{sec:ps} below).  Our final analysis was
thus based on the subset of 80 filaments which are longer than
5\farcm5 ($\sim$ 18 beams).

\begin{figure}
  \centering
\resizebox{\hsize}{!}{\includegraphics[angle=0]{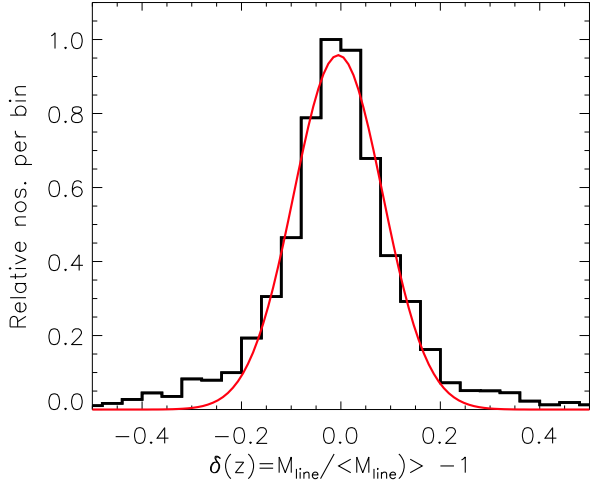}}
\caption{Normalized histogram showing the probability distribution of relative line-mass fluctuations along
the long axes of the subset of 67 filaments which are entirely subcritical.
The red curve shows the Gaussian best fit to the observed distribution, which has a mean of $\sim 0$ and
a standard deviation of $\sim 0.09$.
}
\label{fig:delta_dist}
\end{figure}

\begin{figure}
  \centering
\resizebox{\hsize}{!}{\includegraphics[angle=0]{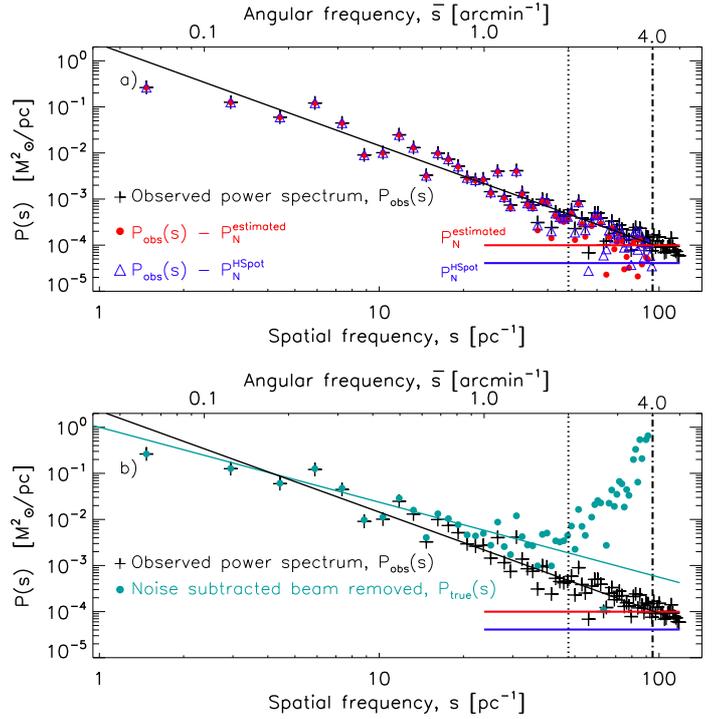}}  
\caption{ Power spectrum of the line-mass fluctuations observed along
  the Pipe filament shown in Fig.~\ref{fig:fluc}, as a function of
  spatial frequency $s$ (bottom x-axis) or angular frequency $\bar{s}$
  (top x-axis).  In both panels, the black plus symbols show the
  observed power spectrum, $P_{\rm obs}(s)$.  In panel a), the red
  dots show the power spectrum, $P_{\rm obs}(s)-P_{\rm N}$, obtained
  after subtracting a white noise power spectrum level $P^{\rm
    estimated}_{\rm N}\sim$ 1$\times 10^{-4}$ \msol$^2$/pc, marked by
  the horizontal red line and estimated from the median value of
  $P_{\rm obs}(s)$ in the 3.9--4.2~arcmin$^{-1}$ angular frequency
  range; the blue triangles show a similar power spectrum after
  subtracting the instrument noise power spectrum level $P^{\rm
    HSpot}_{\rm N}\sim$ 4.0$\times 10^{-5}$ \msol$^2$/pc, marked by
  the horizontal blue line and corresponding to the instrument noise
  level $1\sigma \sim 1\, $MJy/sr in our SPIRE 250~\micron\ maps
  according to HSpot$^3$.
In panel b), the cyan dots show the noise-subtracted {\it and}
beam-corrected power spectrum, $P_{\rm true}(s) = (P_{\rm obs}(s) -
P_{\rm N})/\gamma_{\rm beam}$.  The vertical dotted line in both
panels marks the FWHM of the beam power spectrum at
250~\micron\ ($\bar{s} \approx 2$~arcmin$^{-1}$ -- see
Fig.~\ref{fig:spire-beam}), which is also the highest frequency data
point used to fit a power-law function.  The vertical dot-dashed line
is the Nyquist angular frequency ($\lambda/2D$) for SPIRE
250~\micron\ data.  The power-law fits to the power spectra $P_{\rm
  obs}(s)$ and $P_{\rm true}(s)$ have logarithmic slopes $\alpha_{\rm
  obs}$ = \alphafil\ and $\alpha_{\rm true}$ =
\alphafilbs,\ respectively.  (Considering only angular frequencies up
to $\bar{s}$ $=$ 1.5 arcmin$^{-1}$, the best power-law fit to $P_{\rm
  true}(s)$ has a slope $\alpha_{\rm true} = -1.7\pm0.3$.) }
\label{fig:ps}
\end{figure}

\begin{figure}
  \centering
\includegraphics[scale=0.45]{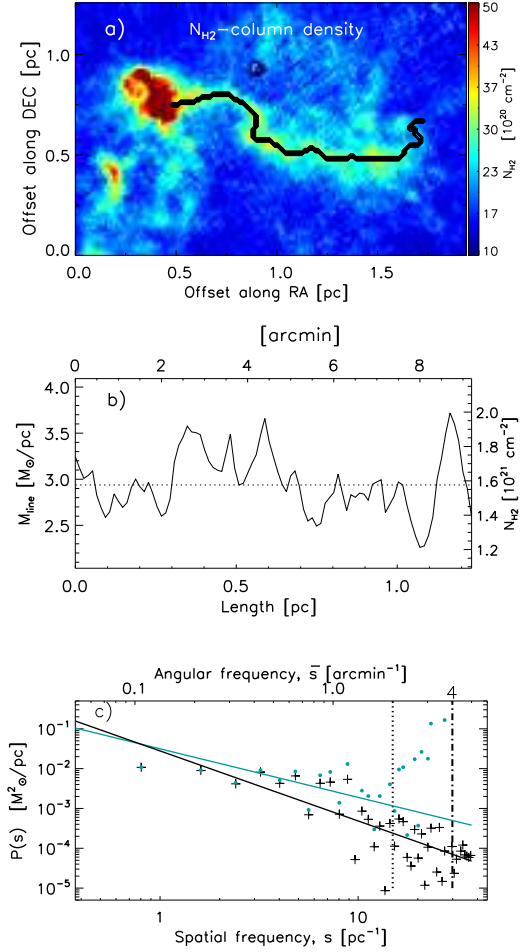}
\caption{Panels ({\bf a}) and ({\bf b}) are similar to
  Fig.~\ref{fig:fluc}a and Fig.~\ref{fig:fluc}b but for an example of
  a thermally subcritical filament in the IC5146 molecular cloud
  (d$\sim$ 460 pc).  The mean line mass of the filament is $<M_{\rm
    line }>$ $\sim$ 3 \msol/pc $<<$ $M_{\rm line,crit}$, as shown by
  the dotted horizontal line in panel ({\bf b}).  The median relative
  amplitude of the line-mass fluctuations is $ \frac{ \lvert M_{\rm
      line}- \langle M_{\rm line} \rangle \rvert}{\langle M_{\rm line}
    \rangle} \sim 0.06$, within the linear perturbation regime.  Panel
  ({\bf c}) is similar to Fig.~\ref{fig:ps} but for the IC5146
  filament.  The power-law fits to the power spectra $P_{\rm obs}(s)$
  and $P_{\rm true}(s)$ have logarithmic slopes $-1.8\pm0.2$ and $-1.2
  \pm 0.4$, respectively.  (Considering only angular frequencies up to
  $\bar{s}$ $=$ 1.5 arcmin$^{-1}$, the best power-law fit to $P_{\rm
    true}(s)$ has a slope $\alpha_{\rm true} = -1.1\pm0.3$.)  }
\label{fig:ic5146}
\end{figure}

\begin{figure}[!h]
  \centering
\includegraphics[scale=.45]{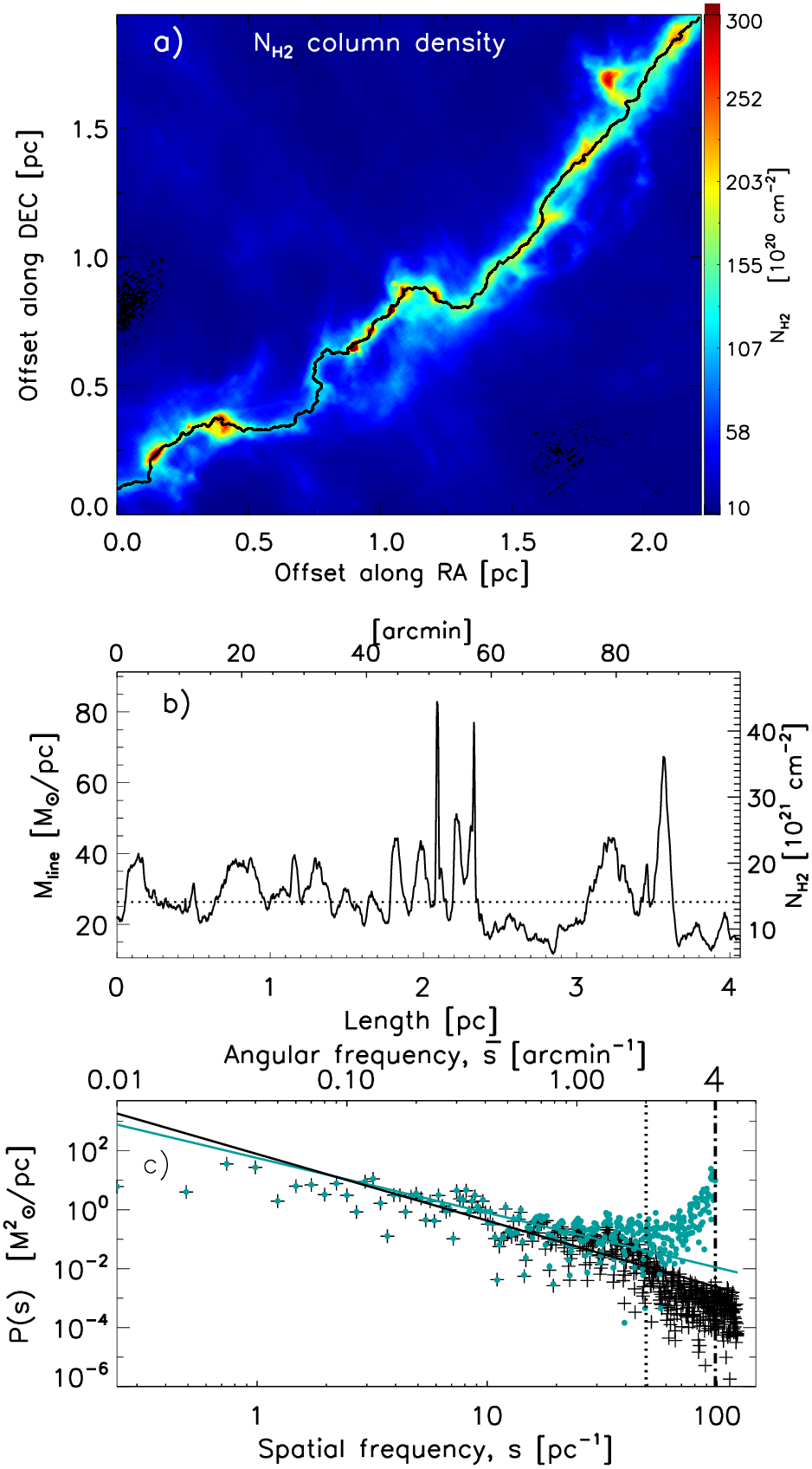}
\caption{Similar to Fig.~\ref{fig:ic5146} but for the thermally
  supercritical filament B211/B213 in the Taurus cloud at d$\sim$ 140
  pc \citep[cf.][]{palmeirim2013}.  The mean line mass of the filament
  is $<M_{\rm line}> \sim 30$ \msol/pc $>>$ $M_{\rm line,crit}$, as
  shown by the dotted horizontal line in panel ({\bf b}).  (Here,
  $<M_{\rm line}>$ was obtained by multiplying the mean
  background-subtracted column density by a characteristic 0.1 pc
  width, whereas \citet{palmeirim2013} obtained a value of 54~\msol/pc
  by integration over the full radial column density profile of the
  filament.)  The median relative amplitude of the line-mass
  fluctuations is $ \frac{ \lvert M_{\rm line}- \langle M_{\rm line}
    \rangle \rvert}{\langle M_{\rm line} \rangle} \sim 0.4$.  Note the
  spike features corresponding to the positions of dense cores along
  the filament and capturing non-linear density perturbations.  In
  panel ({\bf c}), the power-law fits to the power spectra $P_{\rm
    obs}(s)$ and $P_{\rm true}(s)$ have logarithmic slopes
  $-2.2\pm0.2$ and $-1.9 \pm0.3$, respectively.  (The best power-law
  fit to $P_{\rm true}(s)$ is unchanged if only angular frequencies up
  to $\bar{s}$ $=$ 1.5 arcmin$^{-1}$ are considered.) }
\label{fig:taurus}
\end{figure}

\section{Column density/line mass fluctuations along selected filaments} \label{sec:ml}

Figure~\ref{fig:fluc} shows an example of a longitudinal profile of
column density or line-mass fluctuations along a marginally critical
filament in the Pipe molecular cloud.  As the mass per unit length or
line mass is a fundamental variable in cylindrical geometry
\citep[cf.][]{inutsuka2001}, we converted the observed column density
fluctuations to line-mass fluctuations by making use of the
quasi-universal filament width $\sim 0.1$~pc found by
\citet{arzoumanian2011}.  The line-mass fluctuations corresponding to
the example filament displayed in Fig.~\ref{fig:fluc}a are shown in
Fig.~\ref{fig:fluc}b, while the equivalent column density fluctuations
are quantified on the right ordinate axis of Fig.~\ref{fig:fluc}b.  To
estimate fluctuations intrinsic to the selected filaments
{themselves}, a mean line-of-sight background column density was
subtracted from the observed data for each field.  The dotted
horizontal line in Fig.~\ref{fig:fluc}b shows the mean line-mass of
the filament ($\sim$ 15.5 \msol/pc), allowing one to evaluate the
relative strength of the line-mass fluctuations about the mean.  In
this example, as well as for all other subcritical filaments in our
sample, the line-mass fluctuations are small in relative amplitude,
i.e., $ \frac{\lvert M_{\rm line}- \langle M_{\rm line} \rangle \rvert
}{\langle M_{\rm line} \rangle}$ $< 0.1$, implying that the
perturbation modes are in the linear regime (see also
Sect. \ref{sec:discus} below).

Figure~\ref{fig:delta_dist} shows the distribution of normalized
line-mass fluctuations, $\delta =\frac{ M_{\rm line} - <M_{\rm line}
  >}{<M_{\rm line} >} $, along the 67 filaments of our sample which
are subcritical over their entire length.  It can be seen in
Fig.~\ref{fig:delta_dist} that the distribution of line-mass
fluctuations for subcritical filaments is nearly Gaussian, which is an
important assumption of the CMF/IMF model calculations presented by
\citet{inutsuka2001}.  It can also be seen that the density/line-mass
fluctuations along subcritical filaments are typically less than 10\%,
i.e., are in the linear regime.

The power spectrum of the fluctuations observed along the filament
displayed in Fig.~\ref{fig:fluc} is shown in Fig.~\ref{fig:ps} and
discussed in Sect.~\ref{sec:ps} below.  Panels ({\bf a}) and ({\bf b})
of Fig.~\ref{fig:ic5146} and Fig.~\ref{fig:taurus} are similar to
Fig.~\ref{fig:fluc} but show two extreme examples of a subcritical
filament (with $<M_{\rm line }>$ $\sim$ 3 \msol/pc $<<$ $M_{\rm
  line,crit}$) and a supercritical filament (with $<M_{\rm line }>$
$\sim$ 30 \msol/pc $>$ $M_{\rm line,crit}$) in the IC5146 and Taurus
clouds, respectively.

Note that the column density/line mass longitudinal profiles used in
this paper (see, e.g., Figs.~\ref{fig:fluc}b \& \ref{fig:ic5146}b \&
\ref{fig:taurus}b) have an effective 18\farcs2-beam resolution.  The
pixel resolution of the corresponding data is 6\farcs0, close to the
Nyquist sampling limit ($\lambda/2D \sim \, $7\farcs4) of SPIRE
250-\micron\ observations where $D= 3.5$m is the diameter of the
primary mirror of the \her\ telescope.  When deriving physical
properties from the power spectra of individual filaments, we took
special care to exclude angular frequencies higher than the Nyquist
critical frequency $D/\lambda \sim 1/(2 \times 7.4\arcsec) \sim
0.068\, $arcsec$^{-1} \sim 4.0\, $arcmin$^{-1}$.

\begin{figure}[!h]
  \centering
  \includegraphics[scale=.45] {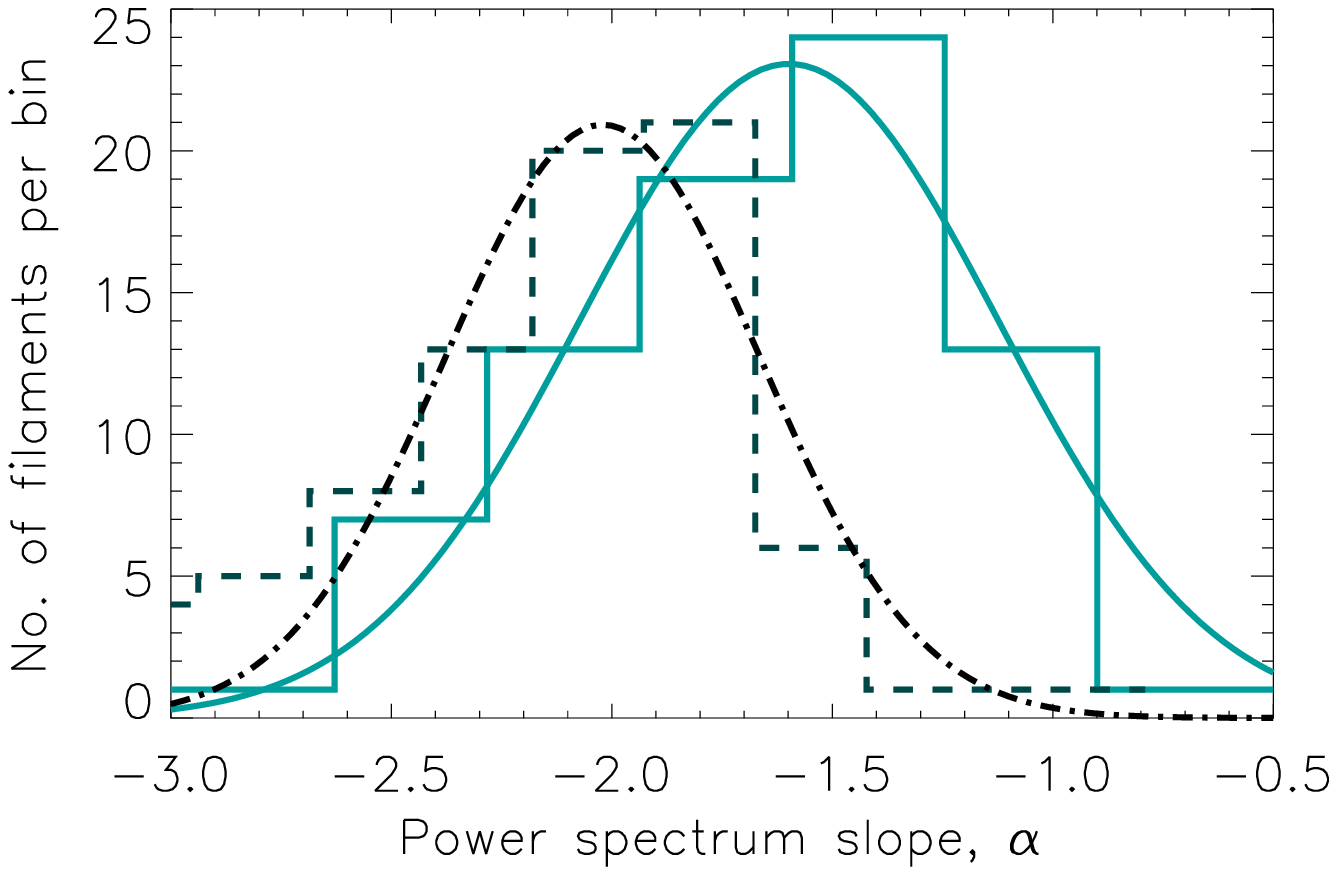}
  \caption{Distributions of power spectrum slopes
  measured before beam correction (dashed histogram) and
  after beam correction (solid histogram)
 for our selected sample of 80 filaments.
   Best-fit Gaussian curves to the two observed distributions are overplotted.
The two distributions are centered on \alphanbs\ and \alphabs\
for the uncorrected and beam-corrected power spectra, respectively.
    }
\label{fig:histo}
\end{figure}

\section{Power spectrum analysis of the filaments}\label{sec:ps}

The normalized power spectrum $P(s)$ of a 1D spatially-varying field $I(l) $ is 
proportional to the square of the Fourier amplitude of the signal, 
and mathematically the relationship in the one-dimensional case is:
\begin{equation}
P(s) = \frac{1}{L} \vert \tilde{I}(s) \vert ^2, 
\label{eqn:ps}
\end{equation}
where $s$ denotes spatial frequency (and $\bar{s}$ denotes angular
frequency), $\tilde{I}(s) = \int \!  I(l) \, e^{-2i\pi s l }\,
\mathrm{d} l $ is the Fourier transform of $ I(l) $, and $L = \int \!
\mathrm{d} l$ is the total length over which the signal is measured.
In the present case, it is convenient to take $I(l) $ to be the field
of line-mass fluctuations (in $M_\odot $/pc) along a given filament
and to express the offset along the filament crest, $l$, in units of
pc.  Then, the normalized power spectrum $P(s) $ has units of
$M_\odot^2 $/pc.

The observed signal after convolution with the telescope beam, $B$, is
given by $I_{\rm obs}({l} )= I_{\rm true}({l}) \ast B + N({l})$, where
$N(l)$ is a noise term arising from small-scale fluctuations primarily
due to instrument noise.  The Fourier transform of $I_{\rm obs}({l})$
is $\tilde{I}_{\rm obs}( {s} )= \tilde{I}_{\rm true }( {s}) \times
\tilde{B}( {s}) + \tilde{N}( {s})$.  Using Eq.~(\ref{eqn:ps}) and the
foregoing relation, it can be seen that the total power spectrum
observed along the filament axis is composed of the true power
spectrum, $P_{\rm true}(s)$, due to line-mass variations along the
filament modified by the power spectrum of the beam, $\gamma_{\rm
  beam}(s)$, plus a white noise power spectrum\footnote{ The SPIRE
  bolometers have a measured temporal stability $\ga 250\, $s, which
  for \her\ parallel-mode observations taken at a scanning speed of
  60\arcsec\ s$^{-1}$ translates to a $1/f$ noise `knee' at spatial
  frequencies corresponding to angular scales larger than $\sim $
  4\degr\ (see Fig. 2 of \citealp{pascale2011}).  The angular scale of
  the longest filament considered here ($\sim$ 2\degr) is
  significantly smaller than both the angular scale of the $1/f$ knee
  and the maximum scan length of the corresponding scan-map
  observations.  We can therefore safely assume that the instrumental
  noise behaves like a white noise over the entire range of spatial
  frequencies considered in the paper.}  term, $P_{\rm N}(s)$, and can
be expressed as:
\begin{equation}
P_{\rm obs}(s)= P_{\rm true}(s) \times \gamma_{\rm beam}(s) +P_{\rm N}(s). 
\end{equation}
In deriving Eq.~(2), we assumed that the noise and the filament
line-mass fluctuations are completely uncorrelated, resulting in a
vanishing cross-power spectrum term.  To estimate the amplitude of
$P_{\rm N}(s)$, we computed the median of the power spectrum in a
narrow band of angular frequencies, 3.9 arcmin$^{-1} \leq \bar{s}
\leq$ 4.2 arcmin$^{-1}$, centered about the Nyquist angular frequency
of 4.0 arcmin$^{-1}$.

In practice, to recover $ P_{\rm true}(s)$, we first computed $P_{\rm
  obs}(s)$ directly from the data, then subtracted a constant noise
power level, and finally corrected the result for the beam convolution
effect by dividing by $ \gamma_{\rm beam}(s)$.
Within the relevant range of angular frequencies ($\bar{s} < $ FWHM of
the beam power spectrum $\sim 2\,$arcmin$^{-1}$), both $P_{\rm
  obs}(s)$ and $P_{\rm true}(s)$ can be approximated and fitted by a
power-law function:
\begin{equation}
P_{\rm obs/true}(s) \approx P(s_0)(s/s_0)^{\alpha_{\rm obs/true}},
\label{eq:psfil}
\end{equation}
where $P(s_0)$ is the normalizing amplitude of the power spectrum at 
spatial frequency $s_0$ and $\alpha$ is the power-law index 
of the fitted power spectrum.

The results of such a power spectrum analysis are shown in
Fig.~\ref{fig:ps} for the marginally critical filament of
Fig.~\ref{fig:fluc}.  The plus symbols in Fig.~\ref{fig:ps} correspond
to the observed power spectrum of the example filament, $P_{\rm
  obs}(s)$, prior to noise subtraction and beam correction.  The upper
panel of Fig.~\ref{fig:ps} explicitly shows the effect of noise level
subtraction.  The horizontal red line marks the estimated power level
due to instrument noise, $P^{\rm estimated}_{\rm N}$ = 1$\times
10^{-4}$ \msol$^2$/pc, derived following the above recipe.  For
comparison, the blue horizontal line shows the noise power spectrum
level ($P^{\rm HSpot}_{\rm N}$) corresponding to the rms instrument
noise level (i.e., $1\sigma \sim 1$~MJy/sr or $\sim 10$ mJy/beam)
expected for SPIRE 250~\micron\ data corresponding to two orthogonal
scans taken at $60\arcsec \rm{s}^{-1}$ in parallel-mode observations
according to the \her\ Observation Planning Tool
(HSpot)\footnote{http://
  herschel.esac.esa.int/Docs/HSPOT/html/hspot-help.html}.
The noise power spectrum level $P^{\rm HSpot}_{\rm N}$ was determined
using Parseval's theorem which relates the variance of the
instrumental noise to the integral of its power spectrum: $\sigma_{\rm
  Mline, HSpot}^2 = \int_{-s_{\rm max}}^{s_{\rm max}} P_{\rm N}^{\rm
  HSpot} ds = 2\, s_{\rm max} \times P_{\rm N}^{\rm HSpot} $, where
the integral is taken over the relevant spatial frequency range given
the maximum angular frequency $ \bar{s}_{\rm max} = 4.0\,
$arcmin$^{-1}$ sampled by our SPIRE 250~\micron\ data.  For a 0.1-pc
wide filament with a median dust temperature of 13.5~K as measured
along the filament crest in Fig.~\ref{fig:fluc}, a $1\sigma $ rms
level of $ \sim 1$~MJy/sr at 250~\micron\ corresponds to a standard
deviation $\sigma_{\rm Mline, HSpot} \sim 0.09\, M_\odot$/pc in
line-mass units.  Given the distance $d = 145$~pc to the filament of
Fig.~\ref{fig:fluc}, this yields $P^{\rm HSpot}_{\rm N} = \sigma_{\rm
  Mline, HSpot}^2 / (2\, s_{\rm max}) \approx 4.0\times 10^{-5}\,
M_\odot ^2$/pc in the same unit as that used on the y-axis of
Fig.~\ref{fig:ps}.

The red dots and overplotted blue triangles in Fig.~\ref{fig:ps}a show
the power spectrum data points after subtracting the estimated noise
spectrum level $P^{\rm estimated}_{\rm N}$ and the expected instrument
noise spectrum level $P^{\rm HSpot}_{\rm N}$, respectively.  It is
apparent from Fig.~\ref{fig:ps}a that the departure of the
noise-subtracted data points from the observed filament power spectrum
is negligible in the angular frequency range $\bar{s} <$
2~arcmin$^{-1}$ where we fit a power-law slope.  Therefore, even
though our estimated noise spectrum level, $P^{\rm estimated}_{\rm
  N}$, likely overestimates the actual noise spectrum level, the step
of noise subtraction has effectively no influence on the estimated
power-law slope of the observed power spectrum.

The beam-corrected power spectrum, $P_{\rm true}(s)$, shown by the
cyan solid dots in Fig.~\ref{fig:ps}b was obtained by dividing the
noise-subtracted power spectrum by the 250~\micron\ SPIRE beam power
spectrum ($\gamma_{\rm beam}$).  The derivation of the SPIRE beam
power spectrum is described in Appendix~\ref{sec:beamps}.  The $
P_{\rm obs}(s) $ spectrum has more power at low spatial frequencies
$s$ (large scales) and decreases roughly as a power law toward high
spatial frequencies until it merges with the noise power spectum level
at a spatial frequency close to $ s \sim$ 100 pc$^{-1}$ (i.e., angular
frequency $\sim$ 4.0 arcmin$^{-1}$).  The beam-corrected power
spectrum, $P_{\rm true}(s)$, has a slightly shallower slope than $
P_{\rm obs}(s) $.  At spatial frequencies higher than the FWHM of the
beam power spectrum shown by the dotted vertical line at $s \sim 50~
$pc$^{-1}$ (angular frequency $\bar{s}\sim$ 2 arcmin$^{-1}$) in
Fig.~\ref{fig:ps}, $P_{\rm true}(s)$ is overcorrected due to
amplification of residual noise when dividing by the beam power
spectrum. This unphysical rise of $P_{\rm true}(s)$ (see
Fig.~\ref{fig:ps}b) at very small angular frequencies is a generic
feature of beam-corrected power spectra and is well documented in
earlier {\it Herschel} studies of cirrus noise
\citep[e.g.][]{martin2010,mamd2010}.  To get around this problem, it
is customary to avoid angular frequencies greater than the FWHM of the
beam power spectrum, i.e., $\bar{s}$ $>$ 2 arcmin$^{-1}$ at 250~$\mu$m
(\citealp[cf.][]{mamd2010}), a compromise that we also adopt in this
paper.  Upon visually inspecting Fig.~\ref{fig:ps}b, there may still
be an indication of a slight over-correction of $P_{\rm true}(s)$ in
the frequency range 1.5~arcmin$^{-1}$~$\lesssim \bar{s}$ $\leq$ 2
arcmin$^{-1}$.  Therefore, we also report the results of more
conservative power-law fits obtained by fitting $P_{\rm true}(s)$ only
up to $\bar{s}$ $=$ 1.5 arcmin$^{-1}$.  Finally, it is common practice
to avoid very small spatial frequencies when fitting power spectrum
slopes in the case of 2D images, due to reasons such as edge effects
or large scanning lengths.  In principle, in the case of a filament
much smaller than the size of the mapped region, there should be no
technical difficulty in reconstructing features at all angular scales
with Fourier modes.  To be on the safe side, we nevertheless excluded
the smallest non-zero angular frequency data point when fitting a
power law to the power spectrum of each filament\footnote{Note that
  including the lowest $s$ data point when fitting the power spectrum
  slope would change the overall distribution of slopes in
  Fig.~\ref{fig:histo} by less than one standard deviation.}.

For the Pipe filament shown in Fig.~\ref{fig:ps}, the beam-corrected
power spectrum has a best-fit slope $\alpha_{\rm
  true}$=\alphafilbs\ and an amplitude $P(s_0=10\rm ~pc^{-1}$)=
\ampfilbs.  For comparison, the observed power spectrum has a somewhat
steeper best-fit slope $\alpha_{\rm obs}$=\alphafil\ and an amplitude
$P(s_0=10 \rm ~pc^{-1})$= \ampfil.  Two other examples of filament
power spectra are shown in Fig.~\ref{fig:ic5146}c and
Fig.~\ref{fig:taurus}c for the IC5146 subcritical filament of
Fig.~\ref{fig:ic5146}a and the Taurus supercritical filament of
Fig.~\ref{fig:taurus}a, respectively.

\begin{figure}[!h]
  \centering
  \includegraphics[scale=.45]{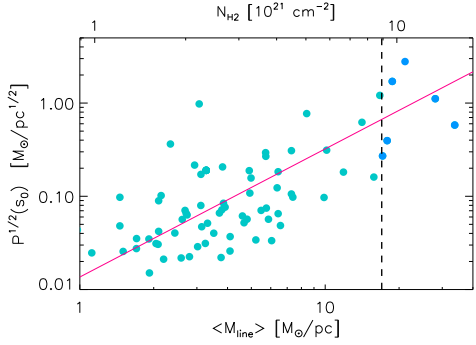} 
  \caption {Square root of filament power spectrum amplitude at $s_0$
    = 10 pc$^{-1}$ versus mean line mass of each filament.  The solid
    line shows the best-fit power-law $P^{1/2}(s_0) \propto M_{\rm
      line}^{1.4\pm0.1}$.  The corresponding mean \nhtwo-column
    density level is shown by the upper coordinate axis.  The vertical
    dashed line separates subcritical filaments (cyan filled circles
    on the left) from supercritical filaments (blue filled circles on
    the right).}
\label{fig:pknh}
\end{figure}

\subsection{A characteristic power spectrum slope for filaments} \label{sec:char_ps}

 To reach statistically significant conclusions on the typical power
 spectrum slope, we considered a sample of 80 subcritical (or
 marginally supercritical) filaments belonging to three distinct
 nearby molecular clouds (see Sect.~\ref{sec:fil_identification}).
 Since the reliability of the derived power spectrum properties also
 depends on the dynamic range covered in spatial frequency space, we
 excluded from our present analysis filaments which were smaller than
 5\farcm5 in the plane of sky.  The resulting distribution of power
 spectrum slopes is shown in Fig.~\ref{fig:histo}.  The two histograms
 shown by the dashed and solid lines represent the distributions
 obtained for the observed [$P_{\rm obs}(s)$] and beam-corrected
 [$P_{\rm true}(s)$] filament power spectra, respectively.  Both
 histograms are well fitted by Gaussian distributions (see
 Fig.~\ref{fig:histo}).

The mean power-law index \alphanbs\ measured for the observed power
spectra is slightly steeper than the mean power-law index
\alphabs\ obtained at $\bar{s}$ $<$ 2 arcmin$^{-1}$ after correcting
for the beam convolution effect.  Fitting $P_{\rm true}(s)$ with a
power law over the more conservative range of angular frequencies
$\bar{s}$ $\leq$ 1.5 arcmin$^{-1}$ yields essentially the same index
within errors, $\bar{\alpha}_{\rm corr} = -1.8\pm0.5$.  The
distribution of beam-corrected power spectrum slopes has a larger
dispersion than the distribution of uncorrected slopes, due to the
propagation of uncertainties resulting from the amplification of
residual noise at small scales.

\begin{figure}[!h]
\centering
\includegraphics[scale=.5]{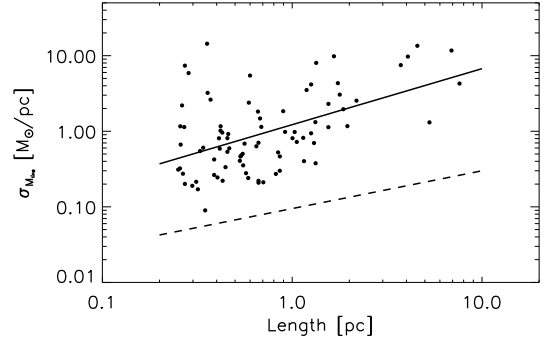}
\caption{ Dispersion of line-mass fluctuations $\sigma_{M_{line}}$
  versus projected length $L$ for the filaments in our sample.  The
  solid line shows the best-fit power-law $\sigma_{M_{line}} \propto
  L^{0.7}$ to the data points.  For comparison, the dashed line shows
  a power-law scaling equivalent to the well-known linewidth--size
  relation in molecular clouds $\sigma_{\rm v} \propto L^{0.5}$
  (cf. \citealp{larson1981}), with an arbitary normalization.}
\label{fig:sig_mline_len}
\end{figure}

\subsection{Correlation between $P(s_0)$ and mean column density}

Figure~\ref{fig:pknh} shows a positive correlation between the
amplitude of the line-mass fluctuations, expressed by $P^{1/2}(s_0)$,
and the mean column density of the filaments.  In other words,
filaments with higher mean column densities tend to have higher power
spectrum amplitudes, i.e., stronger column density fluctuations.  The
red straight line in Fig.~\ref{fig:pknh} shows the best-fit power law
to the data points, which has an index of \pmline. 
Physically, this empirical relation means that the rms fluctuations of
the perturbative modes inside a filament increase with the mean line
mass of the filament.  A similar empirical correlation, with a
somewhat different power-law index ($P_{\rm cirrus} \propto \langle
S_{\nu} \rangle^3$) exists between the 2D power spectrum amplitude of
far-infrared cirrus ($P_{\rm cirrus}$) and the average surface
brightness $ \langle S_{\nu} \rangle$ in far-infrared images of the
sky \citep{gautier1992}.

\subsection{Line-mass dispersion versus filament length}\label{sec:line-sigma}

  Figure~\ref{fig:sig_mline_len} shows that the dispersion of
  line-mass fluctuations along the long axis of each filament,
  $\sigma_{M_{line}}$, is correlated with the physical length $L$ of
  the filament.  The best-fit power-law relation between
  $\sigma_{M_{line}}$ and $ L$ is $\sigma_{M_{line}} \propto L^{0.7}$,
  which is reminiscent of the well-known power-law scaling between
  internal velocity dispersion and region size in molecular clouds
  ($\sigma_{\rm v} \propto L^{0.5}$) originally found by
  \cite{larson1981}.  Such a similarity is not surprising since, in
  the subsonic turbulence regime which approximately holds for
  subcritical or marginally subcritical filaments
  \citep{arzoumanian2013}, density fluctuations are expected to be
  directly proportional to velocity fluctuations.

\section{Concluding remarks}\label{sec:discus} 

Our one-dimensional power spectrum analysis along the axis of 80
nearby {\it Herschel} filaments shows that the longitudinal line-mass
fluctuations along these filaments have a characteristic 1D power
spectrum slope \alphabs.  This result is the first observational
confirmation that the density/line-mass fluctuations along
interstellar filaments have a characteristic power spectrum slope.
The slope we find is remarkably close to the fiducial value
$\alpha_{\rm theory} = -1.5$ adopted by \citet{inutsuka2001} in his
attempt to explain the origin of Salpeter-like power-law tail of the
CMF/IMF at the high-mass end ($dN/dM \propto M^{-2.5}$).  Repeating
Inutsuka's analysis with the observationally-derived power spectrum
slope $\bar{\alpha}_{\rm corr} = -1.6$ found in the present
study\footnote{Here, we adopt the mean, beam-corrected value
  $\bar{\alpha}_{\rm corr} = 1.6$ derived in Sect.~\ref{sec:char_ps}
  as our best estimate of the power spectrum slope. Using the
  prescription of \citet{inutsuka2001} with the mean power-law index
  $\bar{\alpha}_{\rm obs} = -2.0$ measured before beam correction (see
  Fig.~\ref{fig:histo}) instead would lead to a CMF approaching $dN/dM
  \propto M^{-2.2}$ after a few free-fall times.}  would lead to a CMF
approaching $dN/dM \propto M^{-2.4}$ at the high-mass end after a few
free-fall times, which is even closer to a Salpeter mass function
($dN/dM \propto M^{-2.35}$).

Interestingly, the measured power spectrum of density fluctuations along 
interstellar filaments shows a striking similarity to the energy spectrum 
or velocity power spectrum of incompressible hydrodynamical 
turbulence in 1D, i.e.,  
$E(s) \propto P_{\rm v}^{\rm turb}(s) \propto s^{-5/3}$ \citep{kolmogorov1941}\footnote{ Given the uncertainties 
associated with beam correction (see Sect.~\ref{sec:ps}),  the mean beam-corrected index \alphabs\ is also 
marginally consistent with the energy spectrum $E(s) \propto P_{\rm v}^{\rm turb}(s) \propto s^{-2}$ expected 
for  supersonic compressible turbulence \citep[cf.][and references therein]{Padoan2002}.}. 
Note that, as long as self-gravity does not dominate, the continuity equation implies 
a direct proportionality between the Fourier modes of small density fluctuations 
and the modes of subsonic velocity perturbations. 
Our measured power spectrum slope for the density fluctuations (\alphabs) 
is thus consistent with the fact that subcritical (and marginally supercritical) 
filaments, such as the filaments considered in the present study, are 
observed to have subsonic or at most transonic turbulent velocity dispersions 
(\citealp{hacar2011, arzoumanian2013}).

The origin of the characteristic power spectrum slope of filaments
may have a close connection with the process generating 'cirrus' fluctuations 
of far-infrared surface brightness in the interstellar medium.  
The statistical analysis of Galactic `cirrus' fluctuations in $IRAS$ (2D) images  
by \citet{gautier1992} has suggested a plausible role of
interstellar turbulence in shaping these fluctuations.  
The power spectrum of
`cirrus' fluctuations obtained from \cite{roy2010} and recent \her\ images 
\citep{martin2010,mamd2010} has a power-law slope of $-2.7$, 
very similar to the slope of $-8/3 \approx -2.67$ expected for the 
velocity power spectrum of Kolmogorov turbulence in 2D. 
The latter 2D slope translates to a power spectrum slope of $-5/3 \approx -1.67$ in 1D
\footnote{The general relation between the Kolmogorov energy spectrum $E(k) \propto k^{-5/3}$ 
and the velocity power spectrum in $n$-D space is $E(k) \propto k^{n-1}\, P_{\rm v} (k) $.}.

Our {\it Herschel} results support the following speculative picture. 
Filamentary structures form in the cold interstellar medium as a result of a combination 
of 
compression and shear in large-scale magneto-hydrodynamic flows 
\citep{padoan2001,balsara2001,hennebelle2013,inutsuka2015}. 
Supersonic turbulent motions dissipate quickly within dense filaments, forming more stable and coherent structures, 
and leaving an imprint of subsonic 
fluctuations along the filaments. 
Molecular line observations of the inner parts 
of subcritical or marginally supercritical filaments 
do show subsonic or transonic velocity dispersions \citep{hacar2011, arzoumanian2013}, 
supporting the view that the dissipation of turbulence may play a role in the formation 
and evolution of these structures.
The density fluctuations along supercritical filaments are Jeans-unstable and grow 
by local gravitational instability, generating a distribution of prestellar cores.

The observational verification of the existence of a characteristic power spectrum
slope for line-mass fluctuations along interstellar filaments has 
therefore 
opened an interesting avenue for understanding an important aspect of the origin of the prestellar CMF.  
Our findings suggest that an initial spectrum of 
density perturbations seeded by interstellar turbulence may be a prerequisite 
for generating a Salpeter-like CMF at the high-mass end. 
The present study 
combined with previous \her\ results therefore provides insight into 
the possible origin of two distinct features of the CMF:  
the thermal gravitational fragmentation scale of filaments may set the peak of the
lognormal base of the prestellar CMF \citep[e.g.][]{Andre+2014}, while the evolutionary 
characteristics 
of the underlying line-mass perturbations may help to set the 
power-law slope of the CMF/IMF 
at the high mass end (\citealp{inutsuka2001} and this paper). 
A final caveat should be mentioned, however.
There is mounting evidence that true high-mass analogs to low-mass prestellar cores may not exist 
(e.g. \citealp{motte2007}) and that massive protostars may acquire the bulk of their final mass from
much larger scales than a single prestellar core (e.g. \citealp{peretto2013}). 
Further work will be needed to evaluate the relative importance of initial density fluctuations 
and large-scale accretion in generating the high-mass end of the IMF\footnote{Another, more theoretical issue is that, for technical reasons related to the Press-Schechter formalism,  
the model discussed by \citet{inutsuka2001} may, strictly speaking, 
not be a model for the mass function of (prestellar) {\it cores} but a model for the mass function of {\it groups of cores} (P. Hennebelle, private communication).}.

\begin{acknowledgements}
We are thankful to Prof. Shu-Ichiro Inutsuka for stimulating and enlightening discussions on filaments. 
We also acknowledge insightful discussions with Patrick Hennebelle and Gilles Chabrier about the origin of the CMF/IMF 
and the Press-Schechter formalism. 
This work has received support from the European Research Council 
under the European Union's Seventh Framework Programme 
(ERC Advanced Grant Agreement no. 291294 --  'ORISTARS') 
and from the French National Research Agency (Grant no. ANR--11--BS56--0010 -- `STARFICH').
SPIRE has been developed by a consortium of institutes led by
Cardiff Univ. (UK) and including Univ. Lethbridge (Canada);
NAOC (China); CEA, LAM (France); IFSI, Univ. Padua (Italy);
IAC (Spain); Stockholm Observatory (Sweden); Imperial College
London, RAL, UCL-MSSL, UKATC, Univ. Sussex (UK); Caltech, JPL,
NHSC, Univ. Colorado (USA). This development has been supported
by national funding agencies: CSA (Canada); NAOC (China); CEA,
CNES, CNRS (France); ASI (Italy); MCINN (Spain); SNSB (Sweden);
STFC (UK); and NASA (USA).  
PACS has been developed by a consortium of institutes led by MPE
(Germany) and including UVIE (Austria); KUL, CSL, IMEC (Belgium); CEA,
OAMP (France); MPIA (Germany); IFSI, OAP/AOT, OAA/CAISMI, LENS, SISSA
(Italy); IAC (Spain). This development has been supported by the funding
agencies BMVIT (Austria), ESA-PRODEX (Belgium), CEA/CNES (France),
DLR (Germany), ASI (Italy), and CICT/MCT (Spain).
\end{acknowledgements}
\bibliographystyle{aa}
\bibliography{ref,ref_fil}

\begin{thebibliography}{39}
\expandafter\ifx\csname natexlab\endcsname\relax\def\natexlab#1{#1}\fi

\bibitem[{{Alves} {et~al.}(2007){Alves}, {Lombardi}, \& {Lada}}]{alves2007}
{Alves}, J., {Lombardi}, M., \& {Lada}, C.~J. 2007, \aap, 462, L17

\bibitem[{{Andr{\'e}} {et~al.}(2014){Andr{\'e}}, {Di Francesco},
  {Ward-Thompson}, {Inutsuka}, {Pudritz}, \& {Pineda}}]{Andre+2014}
{Andr{\'e}}, P., {Di Francesco}, J., {Ward-Thompson}, D., {et~al.} 2014,
  astro-ph/1312.6232, in Protostars and Planets VI, ed. H. Beuther et al., 27

\bibitem[{{Andr{\'e}} {et~al.}(2010){Andr{\'e}}, {Men'shchikov}, {Bontemps},
  {K{\"o}nyves}, {Motte}, {Schneider}, {Didelon}, {Minier}, {Saraceno},
  {Ward-Thompson}, {di Francesco}, {White}, {Molinari}, {Testi}, {Abergel},
  {Griffin}, {Henning}, {Royer}, {Mer{\'{\i}}n}, {Vavrek}, {Attard},
  {Arzoumanian}, {Wilson}, {Ade}, {Aussel}, {Baluteau}, {Benedettini},
  {Bernard}, {Blommaert}, {Cambr{\'e}sy}, {Cox}, {di Giorgio}, {Hargrave},
  {Hennemann}, {Huang}, {Kirk}, {Krause}, {Launhardt}, {Leeks}, {Le Pennec},
  {Li}, {Martin}, {Maury}, {Olofsson}, {Omont}, {Peretto}, {Pezzuto}, {Prusti},
  {Roussel}, {Russeil}, {Sauvage}, {Sibthorpe}, {Sicilia-Aguilar}, {Spinoglio},
  {Waelkens}, {Woodcraft}, \& {Zavagno}}]{andre2010}
{Andr{\'e}}, P., {Men'shchikov}, A., {Bontemps}, S., {et~al.} 2010, \aap, 518,
  L102

\bibitem[{{Arzoumanian} {et~al.}(2011){Arzoumanian}, {Andr{\'e}}, {Didelon},
  {K{\"o}nyves}, {Schneider}, {Men'shchikov}, {Sousbie}, {Zavagno}, {Bontemps},
  {di Francesco}, {Griffin}, {Hennemann}, {Hill}, {Kirk}, {Martin}, {Minier},
  {Molinari}, {Motte}, {Peretto}, {Pezzuto}, {Spinoglio}, {Ward-Thompson},
  {White}, \& {Wilson}}]{arzoumanian2011}
{Arzoumanian}, D., {Andr{\'e}}, P., {Didelon}, P., {et~al.} 2011, \aap, 529, L6

\bibitem[{{Arzoumanian} {et~al.}(2013){Arzoumanian}, {Andr{\'e}}, {Peretto}, \&
  {K{\"o}nyves}}]{arzoumanian2013}
{Arzoumanian}, D., {Andr{\'e}}, P., {Peretto}, N., \& {K{\"o}nyves}, V. 2013,
  \aap, 553, A119

\bibitem[{{Balsara} {et~al.}(2001){Balsara}, {Ward-Thompson}, \&
  {Crutcher}}]{balsara2001}
{Balsara}, D., {Ward-Thompson}, D., \& {Crutcher}, R.~M. 2001, \mnras, 327, 715

\bibitem[{{Bernard} {et~al.}(2010){Bernard}, {Paradis}, {Marshall}, {Montier},
  {Lagache}, {Paladini}, {Veneziani}, {Brunt}, {Mottram}, {Martin},
  {Ristorcelli}, {Noriega-Crespo}, {Compi{\`e}gne}, {Flagey}, {Anderson},
  {Popescu}, {Tuffs}, {Reach}, {White}, {Benedettini}, {Calzoletti},
  {Digiorgio}, {Faustini}, {Juvela}, {Joblin}, {Joncas}, {Mivilles-Deschenes},
  {Olmi}, {Traficante}, {Piacentini}, {Zavagno}, \& {Molinari}}]{bernard2010}
{Bernard}, J.-P., {Paradis}, D., {Marshall}, D.~J., {et~al.} 2010, \aap, 518,
  L88

\bibitem[{{Chabrier} {et~al.}(2014){Chabrier}, {Hennebelle}, \&
  {Charlot}}]{chabrier2014}
{Chabrier}, G., {Hennebelle}, P., \& {Charlot}, S. 2014, \apj, 796, 75

\bibitem[{{Elias}(1978)}]{elias1978}
{Elias}, J.~H. 1978, \apj, 224, 857

\bibitem[{{Gautier} {et~al.}(1992){Gautier}, {Boulanger}, {Perault}, \&
  {Puget}}]{gautier1992}
{Gautier}, III, T.~N., {Boulanger}, F., {Perault}, M., \& {Puget}, J.~L. 1992,
  \aj, 103, 1313

\bibitem[{{Griffin} {et~al.}(2010){Griffin}, {Abergel}, {Abreu}, {Ade},
  {Andr{\'e}}, {Augueres}, {Babbedge}, {Bae}, {Baillie}, {Baluteau}, {Barlow},
  {Bendo}, {Benielli}, {Bock}, {Bonhomme}, {Brisbin}, {Brockley-Blatt},
  {Caldwell}, {Cara}, {Castro-Rodriguez}, {Cerulli}, {Chanial}, {Chen},
  {Clark}, {Clements}, {Clerc}, {Coker}, {Communal}, {Conversi}, {Cox},
  {Crumb}, {Cunningham}, {Daly}, {Davis}, {de Antoni}, {Delderfield}, {Devin},
  {di Giorgio}, {Didschuns}, {Dohlen}, {Donati}, {Dowell}, {Dowell}, {Duband},
  {Dumaye}, {Emery}, {Ferlet}, {Ferrand}, {Fontignie}, {Fox}, {Franceschini},
  {Frerking}, {Fulton}, {Garcia}, {Gastaud}, {Gear}, {Glenn}, {Goizel},
  {Griffin}, {Grundy}, {Guest}, {Guillemet}, {Hargrave}, {Harwit}, {Hastings},
  {Hatziminaoglou}, {Herman}, {Hinde}, {Hristov}, {Huang}, {Imhof}, {Isaak},
  {Israelsson}, {Ivison}, {Jennings}, {Kiernan}, {King}, {Lange}, {Latter},
  {Laurent}, {Laurent}, {Leeks}, {Lellouch}, {Levenson}, {Li}, {Li},
  {Lilienthal}, {Lim}, {Liu}, {Lu}, {Madden}, {Mainetti}, {Marliani}, {McKay},
  {Mercier}, {Molinari}, {Morris}, {Moseley}, {Mulder}, {Mur}, {Naylor},
  {Nguyen}, {O'Halloran}, {Oliver}, {Olofsson}, {Olofsson}, {Orfei}, {Page},
  {Pain}, {Panuzzo}, {Papageorgiou}, {Parks}, {Parr-Burman}, {Pearce},
  {Pearson}, {P{\'e}rez-Fournon}, {Pinsard}, {Pisano}, {Podosek}, {Pohlen},
  {Polehampton}, {Pouliquen}, {Rigopoulou}, {Rizzo}, {Roseboom}, {Roussel},
  {Rowan-Robinson}, {Rownd}, {Saraceno}, {Sauvage}, {Savage}, {Savini},
  {Sawyer}, {Scharmberg}, {Schmitt}, {Schneider}, {Schulz}, {Schwartz},
  {Shafer}, {Shupe}, {Sibthorpe}, {Sidher}, {Smith}, {Smith}, {Smith},
  {Spencer}, {Stobie}, {Sudiwala}, {Sukhatme}, {Surace}, {Stevens}, {Swinyard},
  {Trichas}, {Tourette}, {Triou}, {Tseng}, {Tucker}, {Turner}, {Vaccari},
  {Valtchanov}, {Vigroux}, {Virique}, {Voellmer}, {Walker}, {Ward}, {Waskett},
  {Weilert}, {Wesson}, {White}, {Whitehouse}, {Wilson}, {Winter}, {Woodcraft},
  {Wright}, {Xu}, {Zavagno}, {Zemcov}, {Zhang}, \& {Zonca}}]{griffin2010}
{Griffin}, M.~J., {Abergel}, A., {Abreu}, A., {et~al.} 2010, \aap, 518, L3

\bibitem[{{Hacar} \& {Tafalla}(2011)}]{hacar2011}
{Hacar}, A. \& {Tafalla}, M. 2011, \aap, 533, A34

\bibitem[{{Heiderman} {et~al.}(2010){Heiderman}, {Evans}, {Allen}, {Huard}, \&
  {Heyer}}]{heiderman2010}
{Heiderman}, A., {Evans}, II, N.~J., {Allen}, L.~E., {Huard}, T., \& {Heyer},
  M. 2010, \apj, 723, 1019

\bibitem[{{Hennebelle}(2013)}]{hennebelle2013}
{Hennebelle}, P. 2013, \aap, 556, A153

\bibitem[{{Hennebelle} \& {Chabrier}(2008)}]{hennebelle2008}
{Hennebelle}, P. \& {Chabrier}, G. 2008, \apj, 684, 395

\bibitem[{{Hildebrand}(1983)}]{Hildebrand1983}
{Hildebrand}, R.~H. 1983, \qjras, 24, 267

\bibitem[{{Hill} {et~al.}(2011){Hill}, {Motte}, {Didelon}, {Bontemps},
  {Minier}, {Hennemann}, {Schneider}, {Andr{\'e}}, {Men'shchikov}, {Anderson},
  {Arzoumanian}, {Bernard}, {di Francesco}, {Elia}, {Giannini}, {Griffin},
  {K{\"o}nyves}, {Kirk}, {Marston}, {Martin}, {Molinari}, {Nguyen Luong},
  {Peretto}, {Pezzuto}, {Roussel}, {Sauvage}, {Sousbie}, {Testi},
  {Ward-Thompson}, {White}, {Wilson}, \& {Zavagno}}]{hill2011}
{Hill}, T., {Motte}, F., {Didelon}, P., {et~al.} 2011, \aap, 533, A94

\bibitem[{{Hopkins}(2012)}]{hopkins2012}
{Hopkins}, P.~F. 2012, \mnras, 423, 2037

\bibitem[{{Inutsuka}(2001)}]{inutsuka2001}
{Inutsuka}, S.-i. 2001, \apjl, 559, L149

\bibitem[{{Inutsuka} {et~al.}(2015){Inutsuka}, {Inoue}, {Iwasaki}, \&
  {Hosokawa}}]{inutsuka2015}
{Inutsuka}, S.-i., {Inoue}, T., {Iwasaki}, K., \& {Hosokawa}, T. 2015, \aap,
  580, A49

\bibitem[{{Inutsuka} \& {Miyama}(1997)}]{inutsuka1997}
{Inutsuka}, S.-i. \& {Miyama}, S.~M. 1997, \apj, 480, 681

\bibitem[{{Jedamzik}(1995)}]{jedamzik1995}
{Jedamzik}, K. 1995, \apj, 448, 1

\bibitem[{{Kolmogorov}(1941)}]{kolmogorov1941}
{Kolmogorov}, A. 1941, Akademiia Nauk SSSR Doklady, 30, 301

\bibitem[{{K{\"o}nyves} {et~al.}(2015){K{\"o}nyves}, {Andr{\'e}},
  {Men'shchikov}, {Palmeirim}, {Arzoumanian}, {Schneider}, {Roy}, {Didelon},
  {Maury}, {Shimajiri}, {di Francesco}, {Bontemps}, {Peretto}, {Motte},
  {Abergel}, {Ali}, {Baluteau}, {Bernard}, {Cambr{\'e}sy}, {Cox}, {di Giorgio},
  {Griffin}, {Hargrave}, {Huang}, {Kirk}, {Li}, {Martin}, {Minier}, {Molinari},
  {Olofsson}, {Pezzuto}, {Russeil}, {Roussel}, {Saraceno}, {Sauvage},
  {Sibthorpe}, {Spinoglio}, {Testi}, {Ward-Thompson}, {White}, {Wilson},
  {Woodcraft}, \& {Zavagno}}]{konyves2015}
{K{\"o}nyves}, V., {Andr{\'e}}, P., {Men'shchikov}, A., {et~al.} 2015, \aap,
  submitted

\bibitem[{{K{\"o}nyves} {et~al.}(2010){K{\"o}nyves}, {Andr{\'e}},
  {Men'shchikov}, {Schneider}, {Arzoumanian}, {Bontemps}, {Attard}, {Motte},
  {Didelon}, {Maury}, {Abergel}, {Ali}, {Baluteau}, {Bernard}, {Cambr{\'e}sy},
  {Cox}, {di Francesco}, {di Giorgio}, {Griffin}, {Hargrave}, {Huang}, {Kirk},
  {Li}, {Martin}, {Minier}, {Molinari}, {Olofsson}, {Pezzuto}, {Russeil},
  {Roussel}, {Saraceno}, {Sauvage}, {Sibthorpe}, {Spinoglio}, {Testi},
  {Ward-Thompson}, {White}, {Wilson}, {Woodcraft}, \& {Zavagno}}]{konyves2010}
{K{\"o}nyves}, V., {Andr{\'e}}, P., {Men'shchikov}, A., {et~al.} 2010, \aap,
  518, L106

\bibitem[{{Lada} {et~al.}(1999){Lada}, {Alves}, \& {Lada}}]{lada1999}
{Lada}, C.~J., {Alves}, J., \& {Lada}, E.~A. 1999, \apj, 512, 250

\bibitem[{{Lada} {et~al.}(2010){Lada}, {Lombardi}, \& {Alves}}]{lada2010}
{Lada}, C.~J., {Lombardi}, M., \& {Alves}, J.~F. 2010, \apj, 724, 687

\bibitem[{{Larson}(1981)}]{larson1981}
{Larson}, R.~B. 1981, \mnras, 194, 809

\bibitem[{{Larson}(1985)}]{larson1985}
{Larson}, R.~B. 1985, \mnras, 214, 379

\bibitem[{{Martin} {et~al.}(2010){Martin}, {Miville-Desch{\^e}nes}, {Roy},
  {Bernard}, {Molinari}, {Billot}, {Brunt}, {Calzoletti}, {Digiorgio}, {Elia},
  {Faustini}, {Joncas}, {Mottram}, {Natoli}, {Noriega-Crespo}, {Paladini},
  {Robitaille}, {Strafella}, {Traficante}, \& {Veneziani}}]{martin2010}
{Martin}, P.~G., {Miville-Desch{\^e}nes}, M.-A., {Roy}, A., {et~al.} 2010,
  \aap, 518, L105

\bibitem[{{Miville-Desch{\^e}nes} {et~al.}(2010){Miville-Desch{\^e}nes},
  {Martin}, {Abergel}, {Bernard}, {Boulanger}, {Lagache}, {Anderson},
  {Andr{\'e}}, {Arab}, {Baluteau}, {Blagrave}, {Bontemps}, {Cohen},
  {Compiegne}, {Cox}, {Dartois}, {Davis}, {Emery}, {Fulton}, {Gry}, {Habart},
  {Huang}, {Joblin}, {Jones}, {Kirk}, {Lim}, {Madden}, {Makiwa}, {Menshchikov},
  {Molinari}, {Moseley}, {Motte}, {Naylor}, {Okumura}, {Pinheiro Gon{\c
  c}alves}, {Polehampton}, {Rod{\'o}n}, {Russeil}, {Saraceno}, {Schneider},
  {Sidher}, {Spencer}, {Swinyard}, {Ward-Thompson}, {White}, \&
  {Zavagno}}]{mamd2010}
{Miville-Desch{\^e}nes}, M.-A., {Martin}, P.~G., {Abergel}, A., {et~al.} 2010,
  \aap, 518, L104

\bibitem[{{Molinari} {et~al.}(2010){Molinari}, {Swinyard}, {Bally}, {Barlow},
  {Bernard}, {Martin}, {Moore}, {Noriega-Crespo}, {Plume}, {Testi}, {Zavagno},
  {Abergel}, {Ali}, {Anderson}, {Andr{\'e}}, {Baluteau}, {Battersby},
  {Beltr{\'a}n}, {Benedettini}, {Billot}, {Blommaert}, {Bontemps}, {Boulanger},
  {Brand}, {Brunt}, {Burton}, {Calzoletti}, {Carey}, {Caselli}, {Cesaroni},
  {Cernicharo}, {Chakrabarti}, {Chrysostomou}, {Cohen}, {Compiegne}, {de
  Bernardis}, {de Gasperis}, {di Giorgio}, {Elia}, {Faustini}, {Flagey},
  {Fukui}, {Fuller}, {Ganga}, {Garcia-Lario}, {Glenn}, {Goldsmith}, {Griffin},
  {Hoare}, {Huang}, {Ikhenaode}, {Joblin}, {Joncas}, {Juvela}, {Kirk},
  {Lagache}, {Li}, {Lim}, {Lord}, {Marengo}, {Marshall}, {Masi}, {Massi},
  {Matsuura}, {Minier}, {Miville-Desch{\^e}nes}, {Montier}, {Morgan}, {Motte},
  {Mottram}, {M{\"u}ller}, {Natoli}, {Neves}, {Olmi}, {Paladini}, {Paradis},
  {Parsons}, {Peretto}, {Pestalozzi}, {Pezzuto}, {Piacentini}, {Piazzo},
  {Polychroni}, {Pomar{\`e}s}, {Popescu}, {Reach}, {Ristorcelli}, {Robitaille},
  {Robitaille}, {Rod{\'o}n}, {Roy}, {Royer}, {Russeil}, {Saraceno}, {Sauvage},
  {Schilke}, {Schisano}, {Schneider}, {Schuller}, {Schulz}, {Sibthorpe},
  {Smith}, {Smith}, {Spinoglio}, {Stamatellos}, {Strafella}, {Stringfellow},
  {Sturm}, {Taylor}, {Thompson}, {Traficante}, {Tuffs}, {Umana}, {Valenziano},
  {Vavrek}, {Veneziani}, {Viti}, {Waelkens}, {Ward-Thompson}, {White},
  {Wilcock}, {Wyrowski}, {Yorke}, \& {Zhang}}]{molinari2010}
{Molinari}, S., {Swinyard}, B., {Bally}, J., {et~al.} 2010, \aap, 518, L100

\bibitem[{{Motte} {et~al.}(1998){Motte}, {Andr{\'e}}, \& {Neri}}]{motte1998}
{Motte}, F., {Andr{\'e}}, P., \& {Neri}, R. 1998, \aap, 336, 150

\bibitem[{{Motte} {et~al.}(2007){Motte}, {Bontemps}, {Schilke}, {Schneider},
  {Menten}, \& {Brogui{\`e}re}}]{motte2007}
{Motte}, F., {Bontemps}, S., {Schilke}, P., {et~al.} 2007, \aap, 476, 1243

\bibitem[{{Offner} {et~al.}(2014){Offner}, {Clark}, {Hennebelle}, {Bastian},
  {Bate}, {Hopkins}, {Moraux}, \& {Whitworth}}]{offner2013}
{Offner}, S.~S.~R., {Clark}, P.~C., {Hennebelle}, P., {et~al.} 2014,
  astro-ph/1312.5326, in Protostars and Planets VI, ed. H. Beuther et al., 53

\bibitem[{{Ostriker}(1964)}]{ostriker1964}
{Ostriker}, J. 1964, \apj, 140, 1056

\bibitem[{{Padoan} {et~al.}(2001){Padoan}, {Juvela}, {Goodman}, \&
  {Nordlund}}]{padoan2001}
{Padoan}, P., {Juvela}, M., {Goodman}, A.~A., \& {Nordlund}, {\AA}. 2001, \apj,
  553, 227

\bibitem[{{Padoan} \& {Nordlund}(2002)}]{Padoan2002}
{Padoan}, P. \& {Nordlund}, {\AA}. 2002, \apj, 576, 870

\bibitem[{{Palmeirim} {et~al.}(2013){Palmeirim}, {Andr{\'e}}, {Kirk},
  {Ward-Thompson}, {Arzoumanian}, {K{\"o}nyves}, {Didelon}, {Schneider},
  {Benedettini}, {Bontemps}, {Di Francesco}, {Elia}, {Griffin}, {Hennemann},
  {Hill}, {Martin}, {Men'shchikov}, {Molinari}, {Motte}, {Nguyen Luong},
  {Nutter}, {Peretto}, {Pezzuto}, {Roy}, {Rygl}, {Spinoglio}, \&
  {White}}]{palmeirim2013}
{Palmeirim}, P., {Andr{\'e}}, P., {Kirk}, J., {et~al.} 2013, \aap, 550, A38

\bibitem[{{Pascale} {et~al.}(2011){Pascale}, {Auld}, {Dariush}, {Dunne},
  {Eales}, {Maddox}, {Panuzzo}, {Pohlen}, {Smith}, {Buttiglione}, {Cava},
  {Clements}, {Cooray}, {Dye}, {de Zotti}, {Fritz}, {Hopwood}, {Ibar},
  {Ivison}, {Jarvis}, {Leeuw}, {L{\'o}pez-Caniego}, {Rigby}, {Rodighiero},
  {Scott}, {Smith}, {Temi}, {Vaccari}, \& {Valtchanov}}]{pascale2011}
{Pascale}, E., {Auld}, R., {Dariush}, A., {et~al.} 2011, \mnras, 415, 911

\bibitem[{{Peretto} {et~al.}(2012){Peretto}, {Andr{\'e}}, {K{\"o}nyves},
  {Schneider}, {Arzoumanian}, {Palmeirim}, {Didelon}, {Attard}, {Bernard}, {Di
  Francesco}, {Elia}, {Hennemann}, {Hill}, {Kirk}, {Men'shchikov}, {Motte},
  {Nguyen Luong}, {Roussel}, {Sousbie}, {Testi}, {Ward-Thompson}, {White}, \&
  {Zavagno}}]{peretto2012}
{Peretto}, N., {Andr{\'e}}, P., {K{\"o}nyves}, V., {et~al.} 2012, \aap, 541,
  A63

\bibitem[{{Peretto} {et~al.}(2013){Peretto}, {Fuller}, {Duarte-Cabral},
  {Avison}, {Hennebelle}, {Pineda}, {Andr{\'e}}, {Bontemps}, {Motte},
  {Schneider}, \& {Molinari}}]{peretto2013}
{Peretto}, N., {Fuller}, G.~A., {Duarte-Cabral}, A., {et~al.} 2013, \aap, 555,
  A112

\bibitem[{{Pilbratt} {et~al.}(2010){Pilbratt}, {Riedinger}, {Passvogel},
  {Crone}, {Doyle}, {Gageur}, {Heras}, {Jewell}, {Metcalfe}, {Ott}, \&
  {Schmidt}}]{pilbratt2010}
{Pilbratt}, G.~L., {Riedinger}, J.~R., {Passvogel}, T., {et~al.} 2010, \aap,
  518, L1

\bibitem[{{Poglitsch} {et~al.}(2010){Poglitsch}, {Waelkens}, {Geis},
  {Feuchtgruber}, {Vandenbussche}, {Rodriguez}, {Krause}, {Renotte}, {van
  Hoof}, {Saraceno}, {Cepa}, {Kerschbaum}, {Agn{\`e}se}, {Ali}, {Altieri},
  {Andreani}, {Augueres}, {Balog}, {Barl}, {Bauer}, {Belbachir}, {Benedettini},
  {Billot}, {Boulade}, {Bischof}, {Blommaert}, {Callut}, {Cara}, {Cerulli},
  {Cesarsky}, {Contursi}, {Creten}, {De Meester}, {Doublier}, {Doumayrou},
  {Duband}, {Exter}, {Genzel}, {Gillis}, {Gr{\"o}zinger}, {Henning},
  {Herreros}, {Huygen}, {Inguscio}, {Jakob}, {Jamar}, {Jean}, {de Jong},
  {Katterloher}, {Kiss}, {Klaas}, {Lemke}, {Lutz}, {Madden}, {Marquet},
  {Martignac}, {Mazy}, {Merken}, {Montfort}, {Morbidelli}, {M{\"u}ller},
  {Nielbock}, {Okumura}, {Orfei}, {Ottensamer}, {Pezzuto}, {Popesso},
  {Putzeys}, {Regibo}, {Reveret}, {Royer}, {Sauvage}, {Schreiber}, {Stegmaier},
  {Schmitt}, {Schubert}, {Sturm}, {Thiel}, {Tofani}, {Vavrek}, {Wetzstein},
  {Wieprecht}, \& {Wiezorrek}}]{poglitsch2010}
{Poglitsch}, A., {Waelkens}, C., {Geis}, N., {et~al.} 2010, \aap, 518, L2

\bibitem[{{Press} \& {Schechter}(1974)}]{press1974}
{Press}, W.~H. \& {Schechter}, P. 1974, \apj, 187, 425

\bibitem[{{Roussel}(2013)}]{roussel2013}
{Roussel}, H. 2013, \pasp, 125, 1126

\bibitem[{{Roy} {et~al.}(2010){Roy}, {Ade}, {Bock}, {Chapin}, {Devlin},
  {Dicker}, {Griffin}, {Gundersen}, {Halpern}, {Hargrave}, {Hughes}, {Klein},
  {Marsden}, {Martin}, {Mauskopf}, {Miville-Desch{\^e}nes}, {Netterfield},
  {Olmi}, {Patanchon}, {Rex}, {Scott}, {Semisch}, {Truch}, {Tucker}, {Tucker},
  {Viero}, \& {Wiebe}}]{roy2010}
{Roy}, A., {Ade}, P.~A.~R., {Bock}, J.~J., {et~al.} 2010, \apj, 708, 1611

\bibitem[{{Roy} {et~al.}(2014){Roy}, {Andr{\'e}}, {Palmeirim}, {Attard},
  {K{\"o}nyves}, {Schneider}, {Peretto}, {Men'shchikov}, {Ward-Thompson},
  {Kirk}, {Griffin}, {Marsh}, {Abergel}, {Arzoumanian}, {Benedettini}, {Hill},
  {Motte}, {Nguyen Luong}, {Pezzuto}, {Rivera-Ingraham}, {Roussel}, {Rygl},
  {Spinoglio}, {Stamatellos}, \& {White}}]{Roy2014}
{Roy}, A., {Andr{\'e}}, P., {Palmeirim}, P., {et~al.} 2014, \aap, 562, A138

\bibitem[{{Sousbie} {et~al.}(2011){Sousbie}, {Pichon}, \&
  {Kawahara}}]{sousbie2011}
{Sousbie}, T., {Pichon}, C., \& {Kawahara}, H. 2011, \mnras, 414, 384

\end{thebibliography}

~
\begin{appendix}
\section{Deriving the power spectrum of the SPIRE 250~\micron\ beam} \label{sec:beamps}

The power spectrum of the SPIRE 250 \micron\ beam was derived from an 
empirical PSF image obtained by the SPIRE ICC from scan map
data\footnote{The reduction of the beam map from scan-map data of Neptune
is described at http://herschel.esac.esa.int/Docs/SPIRE/html/spire\_om.html} 
of Neptune in four directions. There are two sets of empirical PSFs available, one gridded
at the nominal pixel size for each SPIRE passband and the other at a higher pixel 
resolution of 0\farcs6. We used the latter product, but the angular extent of the footprint
was only about 6\farcm0, corresponding to an angular frequency scale of $\sim$ 0.3 arcmin$^{-1}$,
insufficient to correct for the beam convolution effect on large scales (i.e., small $\bar{s}$). 
Therefore, to extend the angular frequency range of the beam power spectrum 
to lower $\bar{s}$  values, we embedded the beam map of Neptune 
inside a larger map 25\farcm0 on a side, and then padded the pixels outside the central Neptune insert 
using the best-fit 2D Gaussian approximation to the beam.

\begin{figure}[!htp]
  \centering
  \resizebox{\hsize}{!}{\includegraphics[angle=0]{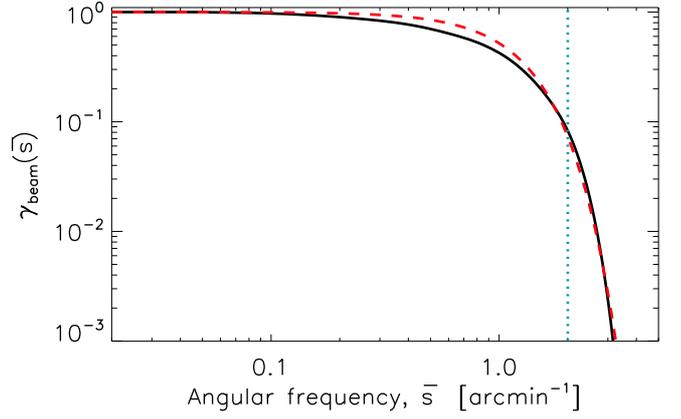}}
  \caption{Power spectrum of the SPIRE 250-\micron\ beam, $\gamma_{\rm beam}(\bar{s})$ (black solid curve). 
The dashed curve represents the power spectrum of a Gaussian beam with a FWHM equal to the nominal 
18\farcs2 HPBW of the SPIRE 250~\micron\ beam.  The  dotted vertical line marks 
the FWHM of the corresponding power spectrum in the angular frequency domain. }
\label{fig:spire-beam}
\end{figure}

The power spectrum of the SPIRE 250~\micron\ beam as a function of angular frequency, $\gamma_{\rm beam}(\bar{s})$,  
is shown in Fig.~\ref{fig:spire-beam}. The power spectrum  
asymptotically converges to unity at small angular frequencies but rapidly declines toward high 
angular frequencies (small angular scales).
We caution that the SPIRE beam spectrum has a slight
suppression of power around 0.2 arcmin$^{-1}$ $<$ $\bar{s}$ $<$ 1.5 arcmin$^{-1}$
compared to a pure Gaussian shape. This problem led
\cite{martin2010} and \cite{mamd2010} to approximate the beam power spectrum
using higher-order polynomial corrections to the Gaussian fit.
The effective full width at half maximum (FWHM) of
the SPIRE 250~\micron\ beam power spectrum\footnote{For a Gaussian beam
the beam power spectrum is also Gaussian, and its FWHM  $\Gamma$,
is related to the half power beam width (HPBW), $\Omega$,
by $\Gamma =\sqrt8\, ln2/\pi\Omega $.} is about 2 arcmin$^{-1}$, as shown by the dashed curve
in Fig.~\ref{fig:spire-beam}.

\section{Results from high-resolution column density maps}

\begin{figure}[!htp]
  \centering
\resizebox{\hsize}{!}{\includegraphics[angle=0]{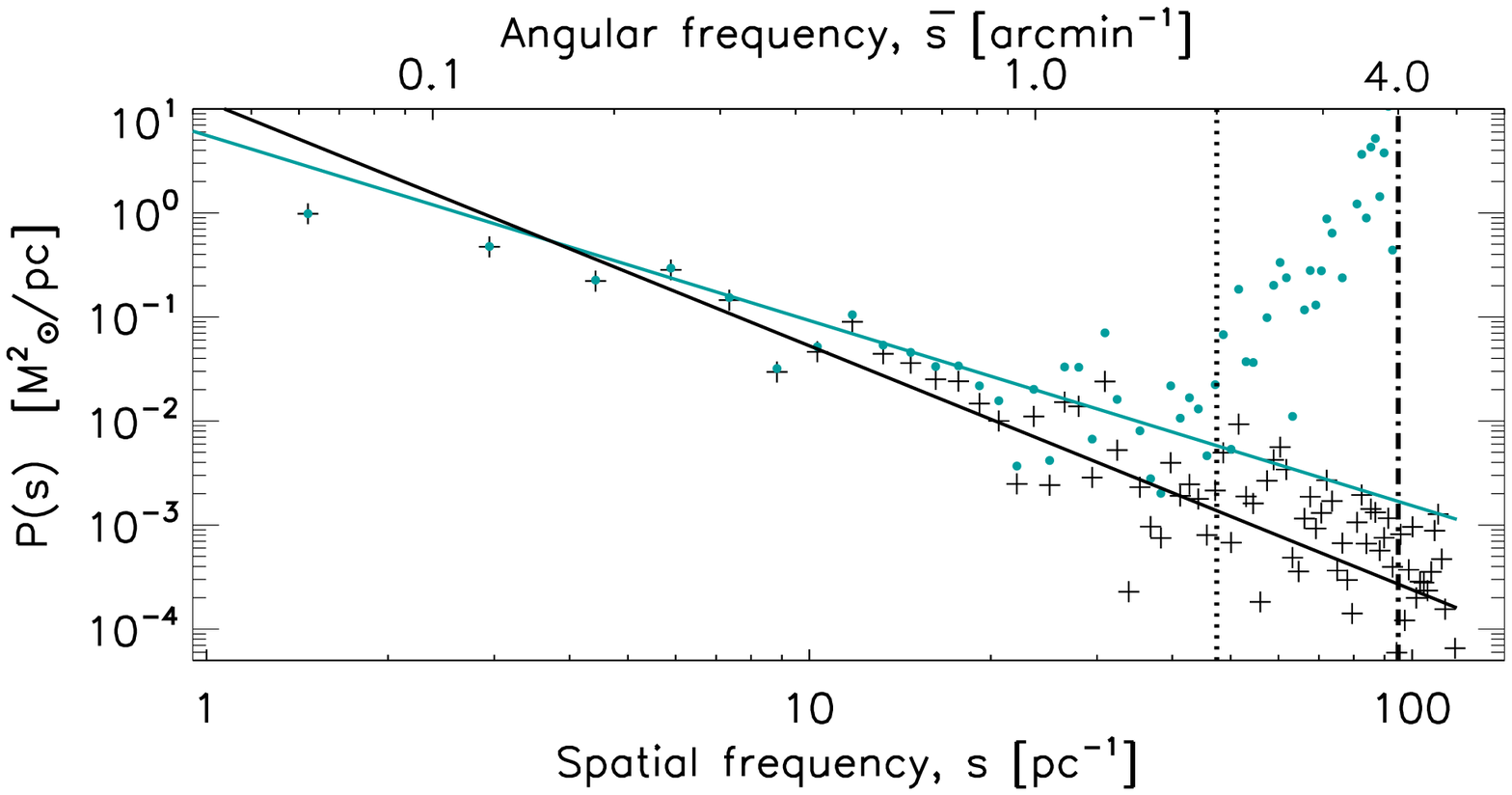}} 
\caption{Power spectrum of the line-mass fluctuations along the Pipe filament of Fig.~\ref{fig:fluc} as measured
in the high-resolution H$_2$-column density map resulting from the multi-scale decomposition technique of \citet{palmeirim2013}.
(Similar to Fig.~\ref{fig:ps} but based on data from the high-resolution column density map instead of the modified 250~\micron\ map.)
The power-law fits to the power spectra $P_{\rm obs}(s)$ (black plus symbols) and $P_{\rm true}(s)$ (cyan filled circles)
have logarithmic slopes $\alpha_{\rm obs} = -2.3\pm 0.2 $ and $\alpha_{\rm corr} =-1.7\pm 0.3$, respectively.
}
\label{fig:A_ps}
\end{figure}

\begin{figure}[htp]
  \centering
\resizebox{\hsize}{!}{\includegraphics[angle=0]{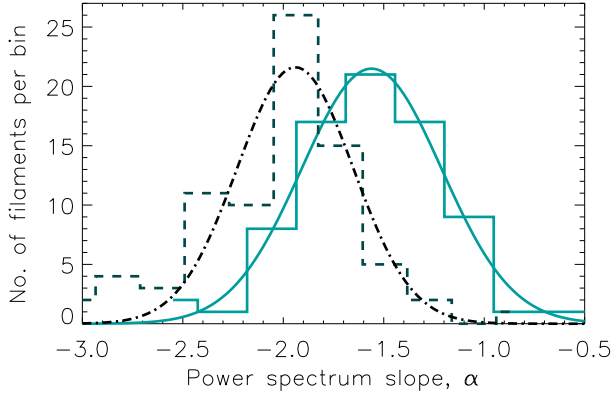}}
  \caption{Distributions of power spectrum slopes
  measured in the high-resolution H$_2$-column density maps
  before beam correction (dashed histogram) and
  after beam correction (solid histogram).
(Similar to Fig.~\ref{fig:histo}
 but based on data from the high-resolution column density maps instead of the modified 250~\micron\ maps.)
Best-fit Gaussian curves to the two observed distributions are overplotted.
The two distributions are centered on $\alpha_{\rm obs}=-1.9 \pm 0.3$ and $\alpha_{\rm corr}=-1.6 \pm 0.45$
for the uncorrected and beam-corrected power spectra, respectively.
    }
\label{fig:A_histo}
\end{figure}

Figure~\ref{fig:A_ps} shows the power spectrum of the same filament already displayed 
in Fig.~\ref{fig:fluc} but derived from a high-resolution column density map 
obtained using the multi-scale decomposition scheme of \citet{palmeirim2013}. 
The power spectrum values at high spatial frequencies has more scatter than
the one derived by converting $I_{\rm 250}$ to H$_2$-column density (see 
Fig.~\ref{fig:ps}).  This is primarily because the multi-scale
decomposition technique introduces fluctuations at small spatial scales which
manifest themselves through enhanced scatter in the Fourier modes at high spatial frequencies.
The distribution of power spectrum slopes derived from high-resolution column density maps 
for the whole sample of filaments 
is shown in Fig.~\ref{fig:A_histo}. The mean power spectrum slopes before and after correcting for
the beam effect are $\bar{\alpha}_{\rm obs}= -1.9\pm0.3$ and $\bar{\alpha}_{\rm corr}= -1.6\pm0.45$, respectively. 
Within the quoted errors, these results are indistinguishable from those found in Sect.~\ref{sec:char_ps} 
using modified 250~$\mu$m maps.

\end{appendix}

\end{document}